\documentclass[%
 reprint,
 amsmath,amssymb,
 aps,
twocolumn
]{revtex4-1}
\usepackage{array,booktabs}
\usepackage{graphicx}% Include figure files
\usepackage{dcolumn}% Align table columns on decimal point

\usepackage{bm}% bold math
\usepackage{float}
\usepackage[table,dvipsnames]{xcolor}% http://ctan.org/pkg/xcolor \usepackage[table, dvipsnames]{xcolor}
\usepackage{caption}
\usepackage{subcaption}

\definecolor{matplotlibblue}{rgb}{0.0, 0.0, 1.0}
\definecolor{matplotlibgreen}{rgb}{0.0, 0.5, 0.0}
\definecolor{matplotlibred}{rgb}{1.0, 0.0, 0.0}
\definecolor{matplotliborange}{HTML}{FF7F0E}
\definecolor{matplotlibmagenta}{rgb}{0.75, 0, 0.75}
\begin{document}

%\preprint{APS/123-QED}

\title{Impact of neutrino pair-production rates in Core-Collapse Supernovae}% 

\author{Aurore Betranhandy}
\affiliation{The Oskar Klein Centre, Department of Astronomy, Stockholm University, AlbaNova, SE-106 91 Stockholm, Sweden\\
}%
\author{Evan O'Connor}%
\affiliation{The Oskar Klein Centre, Department of Astronomy, Stockholm University, AlbaNova, SE-106 91 Stockholm, Sweden\\
}%

\date{\today}

\begin{abstract}
In this paper, we present a careful study on the impact of neutrino pair-production on core-collapse supernovae via spherically-symmetric, general-relativistic simulations of two different massive star progenitors with energy-dependent neutrino transport.  We explore the impact and consequences of both the underlying microphysics and the implementation in the radiation transport algorithms on the supernova evolution, neutrino signal properties, and the explosion dynamics. We consider the two dominant neutrino pair-production processes found in supernovae, electron-positron annihilation as well as nucleon-nucleon bremsstrahlung in combination with both a simplified and a complete treatment of the processes in the radiation transport algorithms.  We find that the use of the simplified prescription quantitatively impacts the neutrino signal at the 10\% level and potentially the supernova dynamics, as we show for the case of a 9.6M$_\odot$ progenitor. We also show that the choice of nucleon-nucleon bremsstrahlung interaction can also have a quantitative impact on the neutrino signal. A self-consistent treatment with state-of-the-art microphysics is suggested for precision simulations of core collapse, however the simplified treatment explored here is both computationally less demanding and results in a qualitatively similar evolution.
\end{abstract}

\maketitle

%\tableofcontents

\section{\label{sec:intro}Introduction}

Core-collapse supernovae (CCSNe) represent the last stage of massive star evolution for stars more massive than 8M$_{\odot}$ and, along with neutron star-neutron star mergers and Type-Ia supernovae, are one of the main channels of galactic nucleosynthesis \cite{woosley_evolution_2002,thielemann_neutron_2017}. Not only do CCSNe contribute to the production of heavy elements but they are also the main birth site of neutron stars and stellar mass black holes. 

Supernovae are also true multimessenger events, producing neutrinos, gravitational waves, as well as photons. The most readily available observable is the electromagnetic signal, for example, the Zwicky Transient Facility observed over 800 CCSNe in 2018 \cite{bruch_large_2020}.  In the fortunate case of a galactic supernova, the two other channels, gravitational waves and neutrinos, become possible \cite{janka_neutrino_2017,kotake_gravitational_2017}. At the onset of the explosion, the outside layers of the progenitor star shroud the core and prevent photons from carrying direct information from the core. Neutrinos and gravitational waves are the only direct channels helping us deciphering the physics of the early explosion.  The supernova mechanism is thus still largely observationally unconstrained. Regardless, numerical simulations performed by different groups allow us to test and refine the theories we have \cite{oconnor_global_2018,mosta_magnetorotational_2014,glas_fast_2019,burrows_three-dimensional_2019,vartanyan_revival_2018,pan_impact_2018,janka_neutrino-driven_2017}. 

The main theory is the neutrino-driven supernova mechanism. Once the fusion reactions in the core stop and gravity overcomes the electron degeneracy pressure, the collapse begins. At nuclear density the core stiffens and the collapse stops and rebounds outwards. The information about this bounce propagates through the in-falling mater, reaching supersonic velocities and creating a shock. This shock propagates out flowing against the ram pressure of the infalling layers and dissociating the nuclei accreting through. In doing so, the shock loses energy and ends up stalling. The neutrino-driven mechanism is the idea \cite{bethe_revival_1985} that the neutrinos can re-energize the shock by transferring energy from the cooling protoneutron star (PNS) to the material behind the shock through absorption in the so-called gain layer. Studies have shown that this heating is very sensitive to the neutrino spectrum, which in turn is sensitive to the emission and absorption processes \cite{bethe_supernova_1990,janka_core-collapse_2012,burrows_perspectives_2013,foglizzo_explosion_2015,janka_neutrino-driven_2017}. While much progress has been made in 3D \cite{vartanyan_successful_2018,vartanyan_temporal_2019,nagakura_towards_2019,murphy_comparison_2019,muller_status_2016,muller_multi-d_2018,muller_three-dimensional_2018,glas_three-dimensional_2019,burrows_three-dimensional_2019,radice_characterizing_2018,powell_gravitational_2018,oconnor_exploring_2018,cabezon_core-collapse_2018}, the theory is not yet completed. Progress on all fronts is needed to further constrain CCSN theories and the underlying physics.  

Neutrino transport is one of the most difficult aspects of modeling supernova simulations. A completely self-consistent treatment of neutrinos would involve solving the 6D Boltzmann equation over the course of the simulation, along with capturing all of the important interactions with the medium.  This is too computationally expensive, especially with the required resolution, though see \cite{nagakura_three-dimensional_2019}. Therefore, we often resort to approximate schemes such as flux limited diffusion or truncated moment schemes. In this study, we use a moment scheme \cite{thorne_relativistic_1981,shibata_truncated_2011}. We essentially treat the neutrinos like a fluid, evolving the energy density and momentum density. Neutrinos are produced inside of the PNS, an optically thick medium, and diffuse out into semi-transparent, optically-thin matter in the gain region behind the shock. This change in the qualitative nature of the environment makes the neutrino transport complex as neutrinos transition from being strongly coupled to the matter to a free-flowing behavior, and therefore neither assumption can be used globally to simplify the problem. The main type of neutrinos interacting in the gain region are the electron type neutrinos and antineutrinos through charged-current interactions. However, heavy lepton neutrinos ($\nu_\mu$, $\bar{\nu}_\mu$, $\nu_\tau$, and $\bar{\nu}_\tau$), which mainly cool the PNS, also play a major role. Hence, their interactions with matter need to be treated as accurately as possible. The main production channel for heavy-lepton neutrinos is via pair-production, where a pair consisting of a neutrino and an antineutrino is formed. The dominate production processes for these pairs in CCSNe include electron-positron annihilation and nucleon-nucleon bremsstrahlung.  Charged-current interactions (either emission or absorption) of single heavy-lepton neutrinos with muons or taus are suppressed due to those charged lepton's large mass, although see \cite{bollig_muon_2017,guo_charged-current_2020,fischer_medium_2020}.

In this paper, different treatments for the thermal pair-production processes are tested. These interactions are challenging to treat as they involve not just one neutrino, but two, which necessitates the coupling of the species and of the energy bins. This had often lead to approximations for their inclusion in neutrino transport algorithms \cite{oconnor_open-source_2015}. As part of this paper, we assess one such approximation with the goal of reducing computational expense while maintaining the fidelity of the solution. Not only are the neutrino pair-production processes computationally complex, another problem inherent in the nucleon-nucleon bremsstrahlung interaction is its uncertain nuclear physics. For this reason, we consider two different ways of treating the interaction. First we consider the commonly used one pion exchange (OPE) formalism  by Hannestad and Raffelt \cite{hannestad_supernova_1998}. We also consider a recent T-matrix formalism formulated by Guo et al. \cite{guo_chiral_2019} based on chiral effective field theory fitted to experimental phase shifts. We test these two different formalisms as well as a simplified version for the nucleon-nucleon bremsstrahlung based on \cite{burrows_neutrino_2006}. For electron-positron annihilation we follow the formalism described in Bruenn et al. \cite{bruenn_stellar_1985} as well as a simplified version \cite{oconnor_open-source_2015}. 
We perform 1D simulations for each of the six combinations of different treatments for two different progenitors. We use a 20-M$_{\odot}$ progenitor, a model studied across many CCSN codes in \cite{oconnor_global_2018} and a 9.6-M$_{\odot}$ progenitor, which has the property of exploding in 1D simulations. We explore the impact of all the different treatments on the early supernova evolution, the explosion parameters, and the neutrino luminosities and mean energies. Furthermore, we scrutinize the validity of our simplified approximation used in order to inform future multidimensional simulations on the impact of heavy-lepton neutrino pair-production treatments.

The outline of this paper is as follows.  In \S~\ref{sec:method}, we overview the simulation code we use, GR1D, and we describe the different interactions involved in our study and their implementation in GR1D and NuLib. We also present the two progenitors and their history of use in CCSN simulations. In \S~\ref{subsec:s20}, we describe the results on the 20M$_{\odot}$ progenitor and \S~\ref{subsec:z9.6} the ones of the 9.6M$_{\odot}$ progenitor. We finally conclude in \S~\ref{sec:conclusion}.

\section{\label{sec:method}Methods}

\subsection{\label{subsec:GR1D}GR1D}

For all the simulations presented in this paper we use the general-relativistic radiation-hydrodynamic code GR1D \cite{oconnor_new_2010,oconnor_open-source_2015}. For the neutrino transport, GR1D uses a moment scheme \cite{shibata_truncated_2011, cardall_conservative_2013}. It evolves the 0th and 1st moment of the neutrino distribution function for multiple neutrino species and multiple neutrino energies. The neutrino-matter interaction terms (completely local) are solved implicitly while the non-local spatial fluxes are solved explicitly. The evolution is done in the coordinate (or laboratory) frame but full velocity dependence is included in the neutrino-matter interactions and to order $v/c$ in the spatial transport terms. We present the model moment evolution equations here, highlighting the neutrino-matter interaction source terms and refer the reader to \cite{oconnor_open-source_2015} for full details,

\begin{equation}
\partial_t [E] + \frac{1}{r^2}\partial_r [\frac{\alpha r^2}{X^2} F_r] + \partial_\epsilon [...] = G_t + C_t 
\end{equation}
and 
\begin{equation}
\partial_t [F_r] + \frac{1}{r^2}\partial_r [\frac{\alpha r^2}{X^2} P_{rr}] + \partial_\epsilon [...] = G_r + C_r
\end{equation}
where $E$ and $F_r$ are the zeroth and first moments of the species and energy-dependent neutrino distribution functions, $P_{rr}$ is the 2nd moment, and in the M1 approximation is taken as an analytic expression involving the first two moments. Here, and in the following, we suppress the energy and species dependence of these moments and source terms, unless needed. $\alpha$ and $X$ are metric functions, $\partial_\epsilon[...]$ refers to the energy-space fluxes, and $G_{t/r}$ and $C_{t/r}$ are the geometric and neutrino-matter source terms, respectively. For the full expression for $\partial_\epsilon[...]$ and $G_{t/r}$ we refer the reader to \cite{oconnor_open-source_2015}, since in this paper we focus on the neutrino-matter interactions, we explicitly write $C_{t/r}$ here and describe each term below,
\begin{equation}
C_t = \alpha^2 [S^t_\mathrm{{e/a}} + S^t_\mathrm{iso} + S^t_\mathrm{scatter} + S^t_\mathrm{pair}] 
\end{equation}
\begin{equation}
C_r = \alpha X^2 [S^r_\mathrm{{e/a}} + S^r_\mathrm{iso} + S^r_\mathrm{scatter} + S^r_\mathrm{pair}] 
\end{equation}

In GR1D, neutrino-matter interactions fall into four categories.  (i) [$S^\alpha_\mathrm{e/a}$]  Charged-current neutrino-matter interactions, where electron type neutrinos and antineutrinos are absorbed or emitted from the matter.   (ii) [$S^\alpha_\mathrm{iso}$] Elastic scattering interactions, where neutrinos of all types scatter on nucleons and nuclei. These scatters change the neutrino direction but maintain their energy. For the emission, absorption and the elastic scattering interactions, we treat the source terms in the following way: 
\begin{align}
    S^{\alpha}_{e/a} &= [ \eta - \kappa_{a} J ] u^{\alpha} - \kappa_{a} H^{\alpha} \label{emission_source_term}\\
    S^{\alpha}_{iso} & = - \kappa_s H^{\alpha} \label{scattering_source_term_eq}
\end{align}
where  $\eta$ is the emissivity, $\kappa_a$ and $\kappa_s$ are the absorption and scattering opacities respectively, $u^\alpha$ is the fluid four-velocity, and $J$ and $H^\alpha$ are the zeroth and first neutrino moments in the fluid frame (see \cite{oconnor_open-source_2015} for detailed expressions of $J$ and $H^\alpha$ in terms of $E$ and $F_r$ and the closure relation).

(iii)[$S^\alpha_\mathrm{scatter}$] Inelastic scattering interactions, where neutrinos scatter on electrons and appreciably change their energy and direction. This interaction necessitates a coupling of neutrino energy bins within a neutrino species. For inelastic neutrino-electron scattering, we use the source terms described in Shibata et al. \cite{shibata_truncated_2011}.   In this study, we ignore inelastic scattering on nucleons.

Finally, (iv) [$S^\alpha_\mathrm{pair}$], pair-production interactions where a neutrino-antineutrino pair is emitted.  In GR1D, we only consider pair-production interactions involving heavy-lepton neutrinos ($\nu_\mu$, $\bar{\nu}_\mu$, $\nu_\tau$, and $\bar{\nu}_\tau$) since the interactions involving electron type neutrino-antineutrino pairs are dwarfed by the charged-current rates for these neutrinos.  With GR1D, there are two ways of including $S^\alpha_\mathrm{pair}$ into the evolution equations. The first is a simplified method where we generate simplified emissivities ($\eta_\mathrm{eff}^{\nu \bar{\nu}}$) and absorption coefficients ($\kappa_{a, \mathrm{eff}^{\nu \bar{\nu}}})$ for each neutrino energy group and treat these terms like the emission and absorption interactions in (i) above. The precise form of these coefficients depends on the particular pair-production process and are described in the following section.  This method is computational efficient as it does not require coupling neutrinos of different energies together when performing the implicit solution of the evolution equations and the interaction rates depend only on the temperature, electron fraction, density and neutrino energy. However, in general, these neutrino pair-production processes do depend on the occupation density of more than one neutrino and therefore this method is an approximation.  The second method is more complete, but also more computationally expensive. It uses kernels to describe the interaction between two neutrinos of different energies (and species) and takes into account the final state neutrino occupation (for emission) and initial state neutrino occupation (for annihilation) hence, coupling different energy groups. The source term for this method is based on \cite{shibata_truncated_2011} and follows from taking the appropriate angular moments of the full Boltzmann collision integral for neutrino-antineutrino annihilation,
\begin{equation}
S^{\alpha}_{\mathrm{pair}}  = \nu^3 \int d\Omega B(\nu,\Omega) (u^\alpha + \ell^\alpha )\,, \label{pp_source_term}
\end{equation}
where $\nu$ is the neutrino energy, $u^\alpha$ is the fluid four-velocity, $\ell^\alpha$ is a unit vector perpendicular to $u^\alpha$, and $B(\nu,\Omega)$ is,
\begin{align}
\nonumber
B(\nu,\Omega) = \int {\nu^\prime}^2  d{\nu ^\prime} d\Omega^\prime [ &( 1 - {f^\prime} ) ( 1 - f) R^{pro}(\nu,\nu^\prime,\mu) \\
&- f {f^\prime} R^{ann}(\nu, \nu^\prime,\mu)]\,,   \label{pp_collision_int} 
\end{align}
where for clarity we have suppressed the $\nu$, and $\nu^\prime$ as well as $\Omega$ and $\Omega^\prime$ dependence in each of the occupation probabilities, $f$ and $f^\prime$, respectively. $\mu$, which is a function of both the prime and unprimed angular variables, is the cosine of the angle between the neutrino and antineutrino.  As is typically done, we assume an angular expansion form of the production and annihilation kernels, $R^\mathrm{pro/ann} \sim R^\mathrm{pro/ann}_0 + \mu R^\mathrm{pro/ann}_1$, where $R_{0/1}^{\mathrm{pro/ann}}$ only depends on the energies of the two neutrinos involved and the underlying interaction (see the following section).  Following \cite{shibata_truncated_2011}, Eqs.~\ref{pp_source_term} and \ref{pp_collision_int} are reduce to a single integral over $\nu^\prime$ where the integrand depends only on the primed and unprimed, zeroth, first, and second neutrino moments and the $R_{0/1}^{\mathrm{pro/ann}}$ kernels,

\begin{align}
\nonumber
S^\alpha_\mathrm{pair} =& \int \frac{d\nu^\prime}{\nu^\prime} \bigg [ -\{(J - 4\pi \nu^3) u^\alpha + H^\alpha\}(4\pi {\nu^\prime}^3 - J^\prime)R_0^\mathrm{pro} \\
\nonumber
&- \frac{{H^\prime}^\alpha}{3}\left\{ (4 \pi \nu^3 - J)R_1^\mathrm{pro} + J R_1^\mathrm{ann}\right\} \\
\nonumber
&+ (h_{\gamma \sigma}H^\gamma {H^\prime}^\sigma u^\alpha + {\tilde L}^\alpha_\beta {H^\prime}^\beta)[R^\mathrm{pro}_1 - R_1^\mathrm{ann}]\\
&- (J u^\alpha + H^\alpha){J^\prime} R_0^\mathrm{ann}\bigg]\,,\label{eq:Spair}
\end{align}
where $h_{\alpha \beta} = g_{\alpha \beta} + u_\alpha u_\beta$ is the projection operator and ${\tilde L}^{\alpha \beta}$ is the traceless $L^{\alpha \beta}$, the second-moment tensor in the fluid frame (analogous to $P^{\alpha \beta}$ above, which is the coordinate frame second moment).

\subsection{\label{subsec:Nulib}Implementation in NuLib}

\begin{table}[htb]
\begin{tabular}{m{4.5cm}p{3.7cm}}
Interaction & Reference\\
\hline
\hline
Emission \& Absorption &\\
\hline
$\nu_e + n$ $\rightleftarrows$  $p  + e^-$ & Bruenn (1985) \cite{bruenn_stellar_1985}; Horowitz (2002) \cite{horowitz_weak_2002}\\
$\bar{\nu}_e + p$ $\rightleftarrows$  $n  + e^+$ & Bruenn (1985) \cite{bruenn_stellar_1985}; Horowitz (2002) \cite{horowitz_weak_2002}\\	
$e^- + A(Z,N)$ $\rightleftarrows$  $A(Z-1,N) + \nu_e$ & Bruenn (1985) \cite{bruenn_stellar_1985}\\[0.3cm]	
Isoenergetic Scattering & \\
\hline
$\nu_i$ + n  $\rightleftarrows$  $\nu_i$ + n & Bruenn (1985) \cite{bruenn_stellar_1985}; Horowitz (2002) \cite{horowitz_weak_2002}\\	
$\nu_i$ + p  $\rightleftarrows$  $\nu_i$ + p & Bruenn (1985) \cite{bruenn_stellar_1985}; Horowitz (2002) \cite{horowitz_weak_2002}\\	
$\nu_i$ + A  $\rightleftarrows$  $\nu_i$ + A & Bruenn (1985) \cite{bruenn_stellar_1985}; Horowitz (1997) \cite{horowitz_neutrino_1997}\\[0.3cm]	
Inelastic Scattering & \\
\hline
$\nu_i$ + e$^-$  $\rightleftarrows$  $\nu_i^\prime$ + $e^{- \prime} $ & Bruenn (1985) \cite{bruenn_stellar_1985}\\[0.3cm]	
Pair Processes & \\
\hline
$e^+ + e^-$  $\rightleftarrows$ $\nu + \bar{\nu}$ & Bruenn (1985) \cite{bruenn_stellar_1985}, Burrows et al. (2006) \cite{burrows_neutrino_2006}, O'Connor (2015) \cite{oconnor_open-source_2015}\\
N + N $\rightleftarrows$ N + N +$\nu + \bar{\nu} $ & Burrows et al. (2006) \cite{burrows_neutrino_2006}, Hannestad and Raffelt (1998) \cite{hannestad_supernova_1998}, \& Guo and Martinez-Pinedo (2019) \cite{guo_chiral_2019}\\[0.3cm]
\end{tabular}
\caption{List of neutrino interactions from NuLib used in this work.}
\label{interaction_table}
\end{table}

NuLib (\url{http://www.nulib.org}) is an open-source neutrino interaction library \cite{oconnor_open-source_2015} that we use to produce tables of the neutrino-matter interaction coefficients for interpolation during our simulations.  For this work, we utilized the interactions described in Table~\ref{interaction_table}, which are divided into the four main interaction types described above. In this work, we focused on the heavy-lepton pair-production processes and the accuracy of the prescriptions used in the transport for these interactions. For this reason, we describe these in detail below. 

The two main neutrino pair-production processes in a CCSN environment are electron-positron pair annihilation and nucleon-nucleon bremsstrahlung. As discussed in \S~\ref{subsec:GR1D}, we consider both an simplified prescription for these interactions and a kernel treatment. For the electron-positron pair annihilation the underlying interaction is the same in these two methods, described in \cite{bruenn_stellar_1985,burrows_neutrino_2006}.  We used NuLib to compute $R_0^\mathrm{ann/pro}$ and $R_1^\mathrm{ann/pro}$ for use in Eq.~\ref{eq:Spair}, which gives the neutrino pair annihilation and production rates as a function of the two neutrino energies, $\nu$ and $\nu^\prime$, for a given value of the matter temperature and electron chemical potential.  For the simplified version of neutrino emission from electron-positron annihilation (see \cite{oconnor_open-source_2015} for more details), we compute $\eta^{e^-e^+}_\mathrm{eff}(\nu)$ by assuming $R_1^{pro/ann}=0$ (i.e. isotropic emission), no final state neutrino blocking, and integrating over all possible $\nu^\prime$.  We construct an simplified absorption by invoking Kirchhoff's law, $\kappa^{e^-e^+}_\mathrm{eff}(\nu) = \eta^{e^-e^+}_\mathrm{eff}(\nu)/BB(\nu,T)$, where $BB$ is the black body intensity for heavy-lepton neutrinos with energy $\nu$ in a medium with temperature $T$. This ensures there is no net emission in regions where the neutrino field is the same as the equilibrium neutrino field and no absorption in regions where the neutrino field is negligible. This is an approximation.  

The other main neutrino pair-production process of importance in CCNSe is nucleon-nucleon bremsstrahlung.  Before this work, this interaction was included in NuLib only via an simplified way taken from Burrows et al. (2006) \cite{burrows_neutrino_2006}.  The simplified single neutrino emissivity (with units of erg cm$^{-3}$ s$^{-1}$ srad$^{-1}$ MeV$^{-1}$) is taken as,
\begin{equation}
\eta^\mathrm{NN}_\mathrm{eff}(\nu) = 0.234 \ \frac{Q_{nb}}{4\pi T} \ \left( \frac{\nu}{T} \right)^{2.4} e^{\frac{-1.1 \nu}{T}}\,,  \label{NNbrem_emi}
\end{equation}
where 
\begin{align}
\nonumber
    Q_{nb}  = &\ 2.0778 \times 10^{30}\  \mathrm{erg}\ \mathrm{cm}^{-3}\  \mathrm{s}^{-1} \times \\
    &\zeta  (x_n^2 + x_p^2 + \frac{28}{3}x_n x_p) \rho_{14}^2 T^{5.5}  \label{Qnb}\,,
\end{align}
is the total energy emission rate for a pair of neutrinos, $\zeta$ is a correction factor (taken to be 0.5 \cite{burrows_neutrino_2006}), $x_{n/p}$ is the mass fraction of neutron and protons, $\rho_{14}$ is the density scaled to 10$^{14}$ g cm$^{-3}$, and $T$ is the matter temperature.  As is the case for electron-positron annihilation, we construct a simplified absorption by invoking Kirchhoff's law, $\kappa^{NN}_\mathrm{eff}(\nu) = \eta^{NN}_\mathrm{eff}(\nu)/BB(\nu,T)$. This simplified emissivity was made in the non-degenerate-medium limit assuming an OPE  potential. It is only dependant on the nucleon number densities and the temperature of the medium. In the early phases of a CCSN explosion, the nucleons at the densities of interest are rarely degenerate, however at latter stages, during the cooling of the PNS for example, the densities where the nucleon-nucleon bremsstrahlung rates can impact the evolution and emission may be in the degenerate regime, therefore this method may need to be reconsidered.\\

In this work, we extend NuLib to include kernels for the nucleon-nucleon bremsstrahlung process in addition to the electron-positron annihilation process.  The nucleon-nucleon bremsstrahlung kernels follow the form of,
\begin{equation}
        R^\mathrm{pro}(\omega,\mu) = G_\mathrm{F}^2 C_a^2 n_B (\hbar c^3) ( 3- \mu) S_\sigma(\omega)\,,\label{source_terme_brems}
\end{equation}

where $n_B$ is the baryon density, $G_\mathrm{F}\sim 1.166\times10^{-11} \mathrm{MeV}^{-2}$ is the weak coupling constant, $C_a=g_A/2$ with $g_A \sim -1.26$ is the axial vector coupling constant, $\omega = \nu + \nu^\prime$ is the sum of the two neutrino energies, and $S_\sigma(\omega)$ is the structure function. As for the electron-positron annihilation kernel, we decompose $R^\mathrm{pro}$ into Legendre moments, $R^\mathrm{pro}_{0/1}$. Given the dependence on $\mu$ in Eq.~\ref{source_terme_brems}, this is a trivial decomposition and $R^\mathrm{pro}_1 = -R^\mathrm{pro}_0/3$. In order to obtain $R_{0/1}^\mathrm{ann}$ in accordance with detailed balance, we use $R_{0/1}^\mathrm{pro}$ = e$^{-\omega/T}$ $R_{0/1}^\mathrm{ann}$. 

The exact definition of $S_\sigma(\omega)$ depends on the underlying interaction and in this work we consider two different models. First, we include the classic nucleon-nucleon bremsstrahlung rates described in Hannestad and Raffelt (1998) \cite{hannestad_supernova_1998}. Similar to the parametrization above, this interaction is derived from the OPE potential, but also includes in the structure function effects such as a non-vanishing pion mass, effects from multiple-scatterings, and is valid for both the degenerate and non-degenerate limits with an interpolation for semi-degenerate regions.  The structure function is \cite{hannestad_supernova_1998},
\begin{equation}
S_\sigma(\omega) =  \frac{\Gamma}{\omega^2 + (\Gamma g(y,\eta) /2 )^2} s(\omega/T,y)\,.  \\
\end{equation}
This structure function is for an arbitrary nucleon interacting with a like nucleon with a nucleon density ($n_N$), temperature ($T$), and the degeneracy factor, $\eta = p_\mathrm{F}^2 / (2m_NT)$ (where $p_\mathrm{F}^2= \hbar (3\pi^2 n_N)^{1/3}$ is the Fermi momentum of the nucleons with mass $m_N$). The spin-fluctuation rate ($\Gamma$), gives the strength of the bremsstrahlung. Also present in the structure function are dimensionless functions $g(y,\eta)$ and $s(\omega/T,y)$ which are a function representing the multi-scattering effect and the interpolation of the nucleon structure function between degenerate and non-degenerate medium, respectively. For completeness, 
\begin{equation}
\begin{split}
    \Gamma =& \frac{8 \sqrt{2\pi} \alpha_{\pi}}{3 \pi^2} \eta^{3/2} \frac{T^2}{m_N c^2}\,, \\
    y =& \frac{m_{\pi}^2}{m_N T}\,,
        \label{OP_terms}
    \end{split}
\end{equation}
where $\alpha_\pi$ and $m_\pi$ are the pion fine-structure constant and the pion mass, respectively. For detailed expressions for $g(y,\eta)$ and $s(\omega/T,y)$, see \cite{hannestad_supernova_1998} or the bremsstrahlung routines in NuLib (\url{http://www.nulib.org}). In NuLib, we compute a table of $R^\mathrm{pro/ann}_{0/1}$ as a function of an arbitrary nucleon density $n_N$, the temperature $T$, for the pair of neutrino energies $\nu$ and $\nu^\prime$.  During our simulation we interpolate this table for three values of the nucleon density, $n_n, n_p, \sqrt{n_nn_p}$, and combine the rates with weights of 1, 1, and 28/3, respectively \cite{burrows_neutrino_2006}.
 
The second nucleon-nucleon bremsstrahlung interaction we consider is the recent formalism from Guo and Martinez-Pinedo \cite{guo_chiral_2019}. They calculate the structure function ($S_\sigma(\omega)$) used in Eq.~\ref{source_terme_brems} by using the T-matrix element based on the $\chi$EFT potential presented in Entem et al. \cite{entem_high-quality_2017}. A similar method was previously explored in Ref.~\cite{bartl_supernova_2014}. A followup in \cite{bartl_impact_2016} where the T-matrix formalism was shown to give modestly different results in supernova simulations from the OPE prescription above. The T-matrix formalism used in \cite{guo_chiral_2019} is an improvement over \cite{bartl_supernova_2014} with the inclusion of off-shell T-matrix elements in addition to on-shell elements.  In NuLib, we utilize the table of $S_\sigma(\omega)$ values provided by the authors. We interpolate this four dimensional table ($\rho$, $T$, $Y_e$, and $\nu+\nu^\prime$) for use in Eq.~\ref{source_terme_brems} in order to construct our tables.

\begin{figure*}
\includegraphics[width=\textwidth]{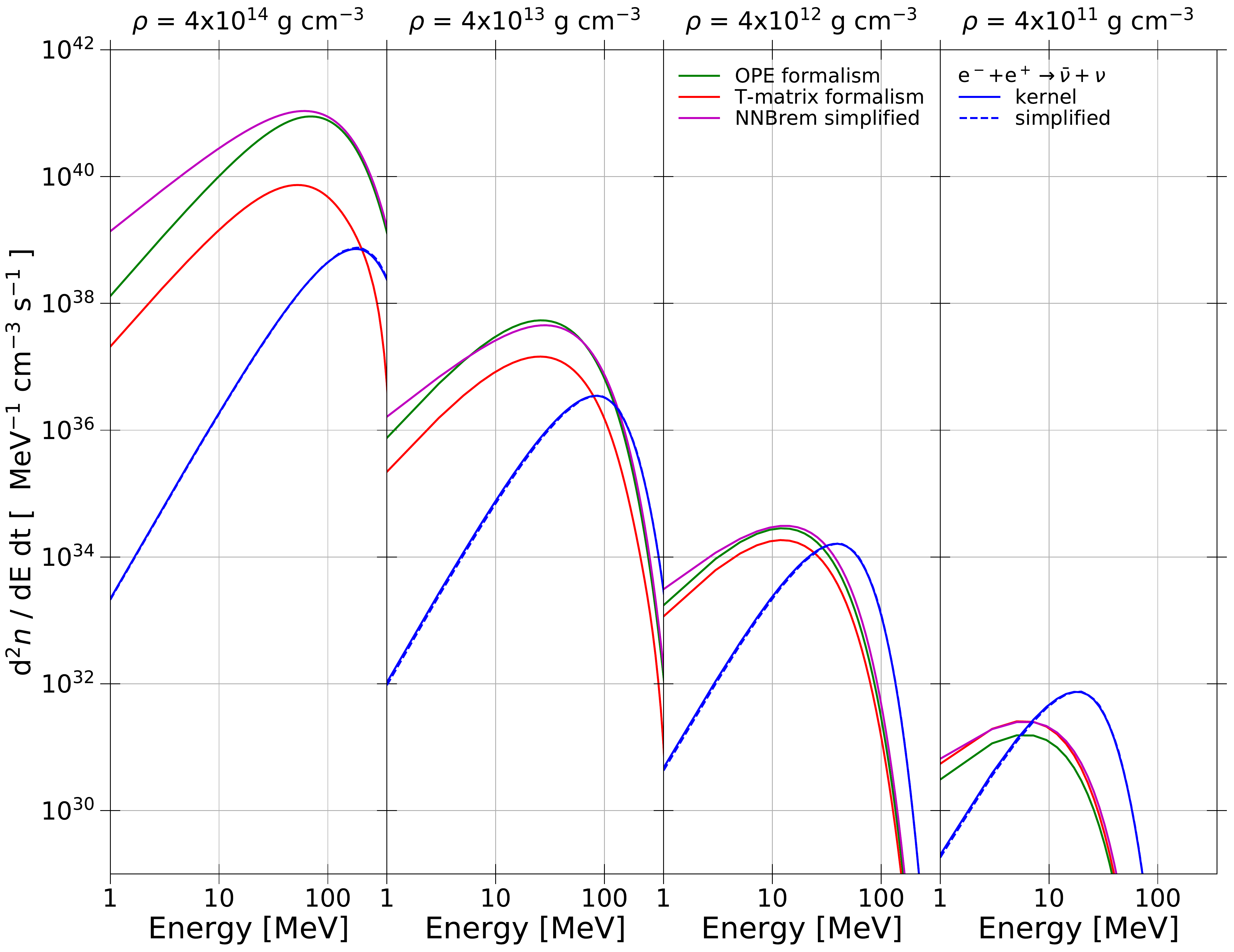}
\caption{Number emissivities for the different pair-production processes for heavy-lepton neutrinos. For the bremsstrahlung we show the emissivity from the Hannestad and Raffelt (1998) \cite{hannestad_supernova_1998} OPE potential kernel (green), the Guo et al (2019) \cite{guo_chiral_2019} T-matrix kernel (red), and the parametrization from Burrows et al. (2006) \cite{burrows_neutrino_2006} (purple). For the electron-positron annihilation we show the emissivity based on the kernels (solid blue) and our parametrization of them (dashed blue), both from Bruenn (1985) \cite{bruenn_stellar_1985}.  We note that for the two electron-positron interactions we expect the same emissivities as the underlying interaction is the same. }
\label{number_production}
\end{figure*}

We conclude this section by comparing each of the pair-production processes and prescriptions utilized in this work at different CCSN-like conditions. The results are shown in Fig.\ref{number_production}, where we compare the single neutrino number isotropic emissivities, ignoring any final state neutrino blocking, as a function of energy at four densities. Following \cite{bartl_supernova_2014}, we use the following relationship between density and temperature typically found in CCSN environments,
\begin{equation}
T_{SN}(\rho) = \text{3 MeV } \left( \frac{\rho}{10^{11} \text{ g cm}^{-3}} \right) ^{\frac{1}{3}}\,,\label{eq:Tsn}
\end{equation}
and adopt an electron fraction of $Y_e =0.2$.\\

The emissivities themselves, as well as the difference between the emissivities, are strongly dependent on the density and temperature. The increase in the bremsstrahlung rates with increasing density is due to both the $\rho^2$ dependence and the roughly $T^{4.5}$ dependence of the number emission rate where the electron-positron annihilation number emission rates increase only to due to the increase in the temperature, scaling roughly as $T^8$. Therefore we expect the importance of bremsstrahlung over electron-positron pair annihilation to scale with the density. Indeed, when the density reaches the typical values of the PNS interior the nucleon-nucleon bremsstrahlung emission dominates. In practice, the core temperatures at densities larger then a few times $10^{13}$\,g\,cm$^{-3}$ do not reach the values predicted from Eq.~\ref{eq:Tsn}, and therefore bremsstrahlung rates dominate over the electron-positron annihilation even more at the highest densities. For the electron-positron pair annihilation, the emissivity derived from the parametrization and the one from the kernel treatment are the same, as expected since the underlying interaction is the same. 

We briefly comment on the differences between the bremsstrahlung treatments. The difference between the T-matrix and OPE treatment is very obvious for densities over 10$^{14}$\,g\,cm$^{-3}$. There, the prescriptions derived from the OPE potential give emissivities more than 10 times greater than the T-matrix prescription. This suppression of the rates at high densities, and also the more modest enhancement of the rates at low density when compared to the OPE interaction is a consequence of the T-matrix treatment \cite{guo_chiral_2019, bartl_supernova_2014}. The parametrization, which is based on the non-degenerate limit of the OPE generally produces comparable rates for the conditions used here. However, we note that the high temperature at nuclear densities resulting from Eq.~\ref{eq:Tsn} are higher then expected during the cooling phase and therefore under those conditions we would expect a larger deviation of the simplified rate from the OPE results. The rates that are expected to be important during the CCSN evolution are the ones near and around the neutrinospheres where the neutrinos are decoupling from the matter.  At high densities, the neutrinos are in equilibrium and the precise rate does not matter, and at low densities the rate is so low that it does not contribute appreciable to the overall neutrino emission. As pointed out in \cite{bartl_impact_2016}, the key densities are around $\rho \gtrsim 10^{12}\ \mathrm{g}\ \mathrm{cm}^{-3}$ during the early core-collapse phase and upward of $\rho \sim 10^{14}\ \mathrm{g}\ \mathrm{cm}^{-3}$ for the cooling phase. Over and above this, it is important to note that the many competing neutrino rates, and their strong temperature dependence, like electron-positron annihilation, often reduce the impact of changes in any one rate.

In addition to the differences that arise from the different interactions (in the case of bremsstrahlung), differences in the actual dynamical evolution can stem from the differences in the transport treatment. As discussed above in \S~\ref{subsec:GR1D}, for the simplified methods, the final state neutrino blocking is not taken in account properly for the emission, nor is the precise form of the annihilation interaction used, rather simplified emission and absorption coefficients are used. With our systematic exploration of these interactions we aim to decipher these differences.

\subsection{\label{subsec:models}Setup}

We performed a set of six simulations with two different progenitors for a total of 12 simulations. A 20-M$_{\odot}$, solar-metallicity, iron-core progenitor \cite{woosley_nucleosynthesis_2007} and a 9.6-M$_{\odot}$ zero-metallicity  iron-core progenitor \cite{heger:pc18} are used. We utilize the 20-M$_{\odot}$ progenitor as it is the same as the one studied in  \cite{oconnor_global_2018} where the evolution was computed using a variety of state-of-the-art evolution codes. The variation done in our study is on the transport treatment of the neutrino pair processes, the remaining physics is held constant. This is an interesting first step to gauge the influence of the different treatments and allows us to quantify the variations against the variations seen between different codes. For this progenitor, we used a grid containing 600 zones with the inner grid spacing being fixed at 300\,m for the inner 20\,km and increasing logarithmically outwards until $\sim1.3\times 10^{10}$\,cm. This progenitor has been explored in many studies, but in particular, Ref.~\cite{just_core-collapse_2018} also consider variations on the neutrino pair-production processes. The other progenitor we consider has a ZAMS mass of 9.6$M_{\odot}$. Unlike most iron-core progenitors, this one has the peculiarity to explode in 1D. Although multidimensional effects can and do impact the development of the explosion in this model \cite{melson_neutrino-driven_2015}, these spherically symmetric simulations give us general insight on the behaviour of the explosion energy development over time and on the neutrino-interaction dependence of the early cooling phase.  For this progenitor,  we used a spherically symmetric grid of 800 zones with a constant grid spacing of 300\,m in the inner 20\,km and then a logarithmically increasing zone size until $\sim 1.3\times 10^9$\,cm. This progenitor has been used in multidimensional studies \cite{mueller_gravitational:13,melson_neutrino-driven_2015, radice_electron-capture_2017}.\\
For all of the simulations, we used the SFHo equation of state from Steiner et al. \cite{steiner_core-collapse_2013} with the same neutrino physics (other than the pair-production treatments) as \cite{oconnor_global_2018}.\\
 The simulation time step is set by the radiation and is equal to the light crossing time of the smallest zone and a CFL condition of 0.4 before bounce, 0.1 near bounce and 0.5 from 20\,ms after bounce for all the simulations. We used a logarithmically spaced energy grid for the neutrinos from 1\,MeV to 250\,MeV with 18 energy groups.

\section{\label{sec:result}Results}
In this section, we explore the impact of the different treatments of heavy-lepton neutrino pair-production described in Sec.\ref{subsec:GR1D} and Sec.\ref{subsec:Nulib} on the supernova evolution. For this, we apply the six different combinations of the pair processes treatments described in Tab.~\ref{run_list}. We will first explore the impact on the 20-M$_{\odot}$ progenitor evolution and follow with the exploding 9.6-M$_{\odot}$ progenitor evolution. 

\begin{table}[htb]
\begin{tabular}{m{0.8cm}m{2.6cm}p{3.7cm}}
\hline
Model & Electron-positron  & Nucleon-Nucleon \\
 & annihilation & Bremsstrahlung\\
 \hline
 \hline
\rowcolor{matplotlibblue!20}
1 & Simplified & Simplified \\
\rowcolor{matplotlibgreen!20}
2 & Simplified & OPE potential formalism\\
\rowcolor{matplotlibred!20}
3 & Simplified & T-matrix formalism \\
\rowcolor{matplotlibblue!80}
4 & Kernel formalism & Simplified \\
\rowcolor{matplotlibgreen!80}
5 & Kernel formalism & OPE potential formalism\\
\rowcolor{matplotlibred!80}
6 & Kernel formalism & T-matrix formalism\\
\end{tabular}
\caption{Enumeration of the different neutrino treatment combinations. Colors are used throughout the figures where light colors are shown via dashed lines. 
}
\label{run_list}
\end{table}

\subsection{\label{subsec:s20}s20 progenitor}

\begin{figure}
\includegraphics[width=0.5\textwidth]{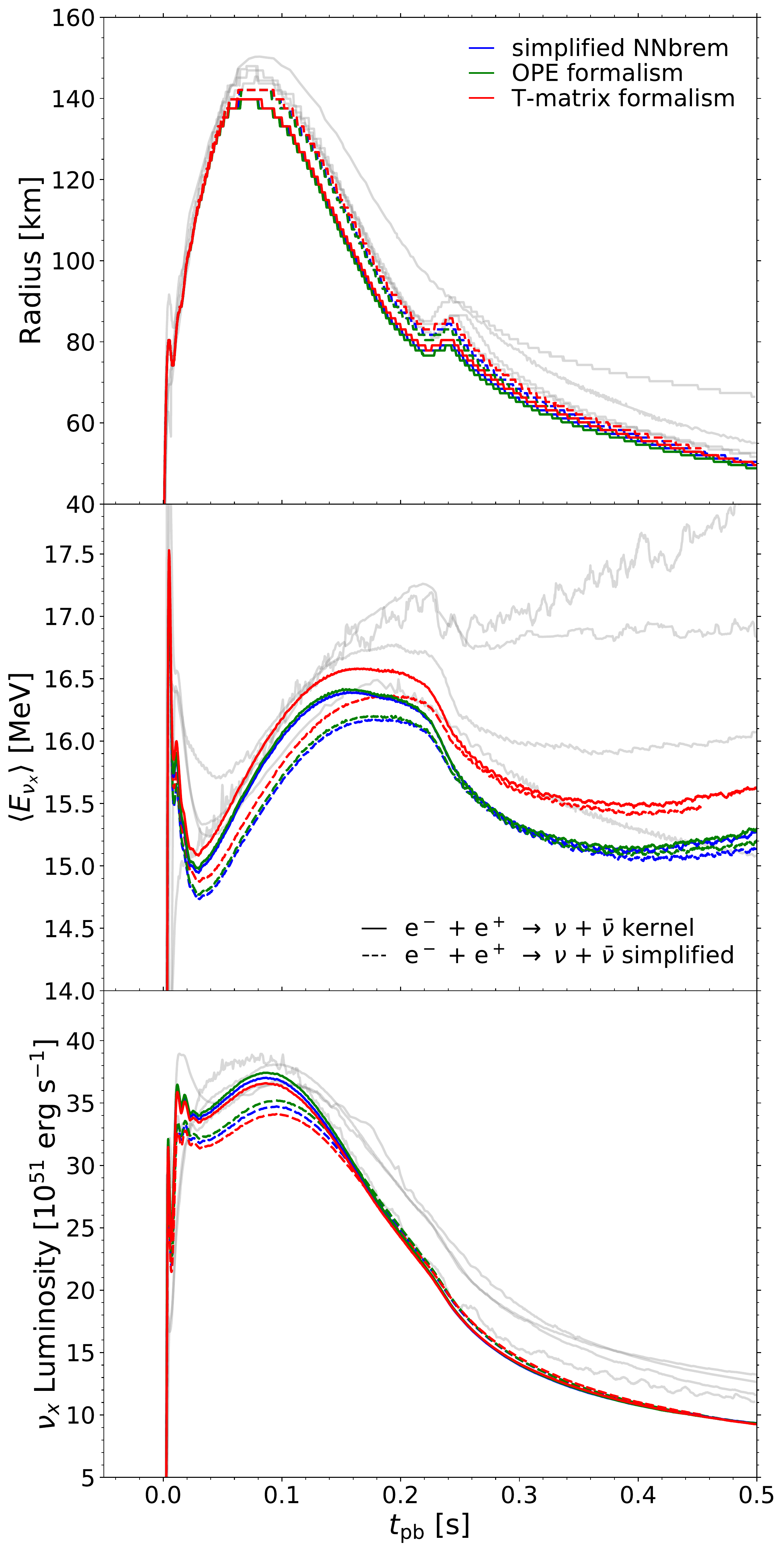}
\caption{Evolution quantities for the 6 models using the 20-M$_{\odot}$ progenitor. In the top panel we show the shock radius evolution, the middle panel shows the $\nu_x$ mean energy, and the bottom panel shows the evolution of the $\nu_x$ luminosity.
}
\label{s20_luminosities}
\end{figure}

The 20-M$_{\odot}$ progenitor does not lead to an explosion.  The shock radius evolutions are plotted in the top panel of Fig.~\ref{s20_luminosities}. The different colors correspond to the different models in Tab.\ref{run_list}. The blue, green, and red solid lines refer to the three simulations using the electron-positron annihilation kernels with the bremsstrahlung fit, OPE kernel, and T-matrix kernel, respectively, while the three dashed lines refer to the electron-positron annihilation simplified emissivity for the three different bremsstrahlung treatments. All of the models give qualitatively similar results. Bounce occurs at $\sim$298\,ms after the onset of collapse. The shock then expands for $\sim$90\,ms after bounce and reaches a radius of $\sim$150\,km where it stalls for $\sim$10\,ms and starts to recede. The shock radius shows a short expansion phase again at $\sim$230\,ms after bounce, which is due to the silicon-oxygen shell interface accreting through the shock front. The shock radius then continues to recede to attain $\sim$50\,km at 500\,ms after bounce. For reference, we show with grey lines the shock radius evolution from simulations with various codes for the same progenitor and setup taken from the comparison study of \cite{oconnor_global_2018}.  The shock evolution of all our models generally agree with these simulations and the level of variation between our simulations is slightly less than that observed between the simulation codes. 

The different neutrino-pair production treatments only modestly impact the shock radius evolution.  For the simulations using the full kernel treatment for the electron-positron-annihilation to neutrino-pair process (models 4, 5, and 6) there is a consistently lower shock radius ($\sim$5\,km) compared to the models with the simplified emissivity for this process (models 1,  2, and 3). This hierarchy is correlated with the properties of the heavy-lepton neutrino emission (bottom panels of  Fig.\ref{s20_luminosities}). As we discuss below, during the first $\sim$150\,ms after bounce, the largest heavy-lepton neutrino luminosities and mean energies arise from the simulations using the full kernel treatment of the electron-positron annihilation pair-production process.  These simulations give enhanced cooling, smaller PNS radii, and a smaller shock radii.  This cause-and-effect is commonly seen in this model (and for which the explosion properties in multidimensional simulations of this model are particularly sensitive too), for example with modifications on the neutral-current scattering opacities \cite{melson_neutrino-driven_2015-1,oconnor_core-collapse_2017}.

In the bottom two panels of Fig.\ref{s20_luminosities}, we show the heavy-lepton neutrino mean energy (middle panel) and the heavy-lepton neutrino luminosity (bottom panel) as measured in the coordinate frame at 500\,km.  For the luminosities, after a small peak at bounce, there is a short rise to a plateau around $\sim 35 \times 10^{51}$\,erg\,s$^{-1}$ at the time of the peak shock radius. The heavy-lepton luminosities then decrease as the PNS contracts reaching values $\sim 10^{52}$\,erg\,s$^{-1}$ at 500\,ms after bounce. For the heavy-lepton neutrino mean energies, after a short peak at bounce, the mean neutrino energy rises from $\sim$30\,ms after bounce from $\sim14.5-15$\,MeV to a peak of $\sim$16-16.5\,MeV. With the accretion of the silicon/oxygen interface the heavy-lepton neutrino mean energy drops $\sim$ 1\,MeV and generally plateaus at $\sim$15-15.5\,MeV until the end of the simulation at 500\,ms. As we have shown for the shock radius, we show the neutrino luminosities and mean energies from \cite{oconnor_global_2018} in grey.  We can see that the different neutrino pair-production formalisms create differences which are comparable to the variability seen across different transport methods and hydrodynamics.  It is worth noting that in \cite{oconnor_global_2018}, the prescriptions of the treatment of heavy-lepton neutrinos also varied among the codes. The impact of the different pair-production treatments on the electron-type neutrino luminosities and mean energies (not shown) is small. 

During all stages of the evolution the quantities in Fig.~\ref{s20_luminosities} are within $\sim$10\% of each other for the luminosities and within $\sim$3\% for mean energies. However, the differences seen do correlate with the different pair-production treatments.  Models 4, 5, and 6, where we use the full kernel-based treatment for the electron-positron annihilation process, have the largest neutrino luminosities and mean energies during the first $\sim$150\,ms after bounce, while the simplified electron-positron annihilation treatment (models 1, 2, and 3) shows consistently lower luminosities and energies during this time. As we mentioned above, this causes increased PNS contraction and lower shock radii for the former models. However, also as a consequence of the increased contraction, there is increased electron neutrino mean energies, and an increased specific neutrino heating (although less overall heating due to the smaller gain region). For all models, the luminosity differences mostly disappear starting at $\sim$200\,ms after bounce, although some differences in the mean energy remain as we will discuss below.  

The impact of the nucleon-nucleon bremsstrahlung treatment on the evolution is less obvious. We do observe that among the different bremsstrahlung treatment, the use of T-matrix formalism  (models 3 and 6) systematically creates a higher neutrino mean energy throughout the entire simulation, but especially after $\sim$250 ms after bounce. This is due to the lower emissivity of this interaction at higher densities (see Fig.~\ref{number_production}) which gives an earlier decoupling radius and therefore a harder spectrum, since the matter temperatures are higher. The luminosities also tend to be the lower soon after bounce when using this formalism.  The differences between the use of the OPE potential kernel-based formalism and the simplified emissivity based on this same potential only appear in the luminosities, and even there it is minimal.  It has the effect of reducing the luminosity for $\sim$160\,ms following bounce, analogous to the use of the simplified emissivity for the electron-positron annihilation, but smaller in magnitude. 

From these observations we conclude that the differences created by the use of the simplified emissivities mainly lie in simplistic treatment of the neutrino transport (i.e. ignoring the functional form of the neutrino and antineutrino distributions and their angular dependence as well as any final state blocking, as explained in \S~\ref{subsec:GR1D}) rather than differences in the underlying neutrino interaction model. A previous study, \cite{just_core-collapse_2018}, explored the impact of a simplified heavy-lepton neutrino pair-production treatments as well. They find similar changes on the luminosity, mean energy and shock radius evolution as the ones we find comparing models 1 and 5. They suggest that the differences seen are a result of the implicit assumption of the angular dependence (i.e. that it is isotropic) of the neutrino annihilation partner, rather than the in situ distribution, which is forward peaked (we note we from Eq.~\ref{source_terme_brems} that the annihilation strength is minimal for co-travelling neutrinos). This over-predicts neutrino-antineutrino annihilations within the simplified emissivity assumption.  While this is certainly true, we note that since the neutrino annihilations are occurring well below the scattering surface, the distribution function is very isotropic. We therefore suggest it is rather the overall magnitude of the occupation density of the annihilation partner (which is implicitly assumed to be the black body distribution) that causes the simplified emissivity to over-predict annihilation and thus lead to smaller emergent heavy-lepton neutrino luminosities. We show this in the following.

\begin{figure*}
    \centering
    \includegraphics[width=\textwidth]{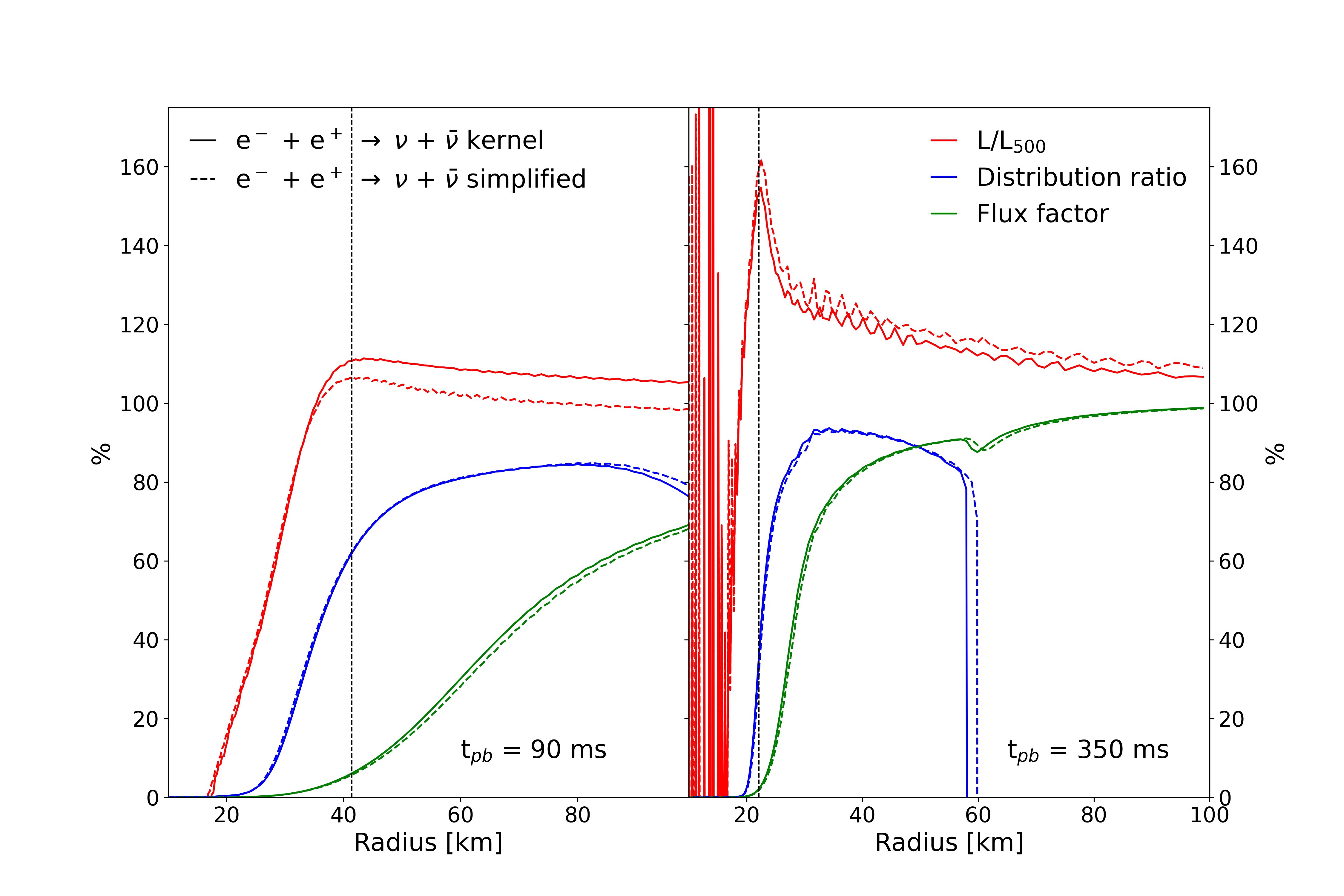}
    \caption{Radial neutrino properties at early (left) and late (right) times for the 20-M$_\odot$ progenitor.  We show results for the kernel treatment of electron-positron annihilation with solid lines and the simplified electron-positron annihilation treatment with dashed lines.  In both cases, we use the T-matrix kernel treatment for the bremsstrahlung interaction. In red we show the radial evolution of the total outgoing neutrino luminosity, normalized to the value to the kernel treatment at 500\,km.  In blue we show the relative difference between the actual and equilibrium neutrino distribution function, while in green we show the flux factor.  For the former radial profile we select an energy bin with $\sim$15\,MeV. Vertical lines denote the peak of the neutrino luminosity for the simplified treatment. 
    }
    \label{flux_factor}
\end{figure*}

 For the 20-$M_\odot$ progenitor at both 90\,ms (left) and 350\,ms (right), we show several key heavy-lepton neutrino properties from both model 3 (simplified electron-positron treatment; dashed line) and model 6 (kernel electron-positron treatment; solid line).  In red, we show the growing outward going heavy-lepton neutrino luminosity, i.e. $4\pi r^2 F_r$, normalized so that the luminosity from the full kernel treatment is 100\% at 500\,km.   In blue, again for model 3 and 6 with dashed and solid line, respectively, we show the difference between the equilibrium heavy-lepton neutrino distribution and the actual heavy-lepton neutrino distribution relative to the heavy-lepton equilibrium distribution, i.e. $(f_\mathrm{eq} - f) / f_\mathrm{eq} = 1-f/f_\mathrm{eq}$ for an energy bin corresponding to $\sim$15\,MeV.  Here we take $f=J/(4\pi \nu^3)$ and $f_\mathrm{eq} = 1/(exp(\nu/T)+1)$.  In green, we show the energy averaged heavy-lepton flux factor (= $F_r/E$).  This allows us to highlight the proportional importance of the anisotropy of the neutrino field (the green lines) and the deviation of the actual neutrino distribution from equilibrium (blue lines). At the early time (left panel), at the radii where the neutrinos luminosity is rising ($\sim$20-40\,km) the distributions are almost isotropic ($F_r/E \sim 15\%$). However at these same radii, the occupation density of the neutrinos significantly deviates from the black body, falling short of the equilibrium occupation density by $\sim60\%$ at $\sim$40\,km. Since the simplified treatment implicitly assume a black body distribution for the annihilation partners, this leads to an over prediction of the annihilation rate in this regime. The result is a further out decoupling radius, and ultimately a lower heavy-lepton neutrino luminosity and mean energy.  This implies that a significant factor in the difference between a full and simplified treatment is linked to the assumed distribution of the pair neutrino rather than the intrinsic anisotropy of the radiation field. 

For late times (right panel), at the radii where the luminosity is rising ($\sim$16-22\,km), the distribution is almost completely isotropic (rising to $F_r/E \sim 5\%$ at 22\,km) and the deviation of the distribution function from the equilibrium distribution is negligible at all radii except for the last 20\% of the emission, even there, the deviation is at most $\sim$30\%. As a result we do not see any excess annihilation at these late times.  It is worth noting that the simplified treatment does actually predict a larger luminosity at these late times, by a few percent.  We suspect this is due to lack of final state blocking in the emission of the neutrinos within the simplified treatment.  At late times the value of the heavy-lepton neutrino distribution function near the peak of the emission is several times the value seen at earlier times, raising the impact of the final state blocking. Some oscillations appear in the luminosity for late times. This is a consequence of the explicit-in-time, first-order, forward-Euler scheme used for the evaluation of the spatial fluxes.  It tends to occur for energy groups that are close to free streaming (with the characteristic speeds entering the Riemann problem approaching $c$) in zones with a CFL condition $c \Delta t / \Delta x \sim 0.5$. It causes the small oscillations seen near $\sim$30\,km and the instabilities in the outward going neutrino flux under 15\,km, which at this point is constrained to the lowest energy groups as the rest are fully trapped.

\subsection{\label{subsec:z9.6}z9.6 progenitor}

\begin{figure}
\includegraphics[width=0.48\textwidth]{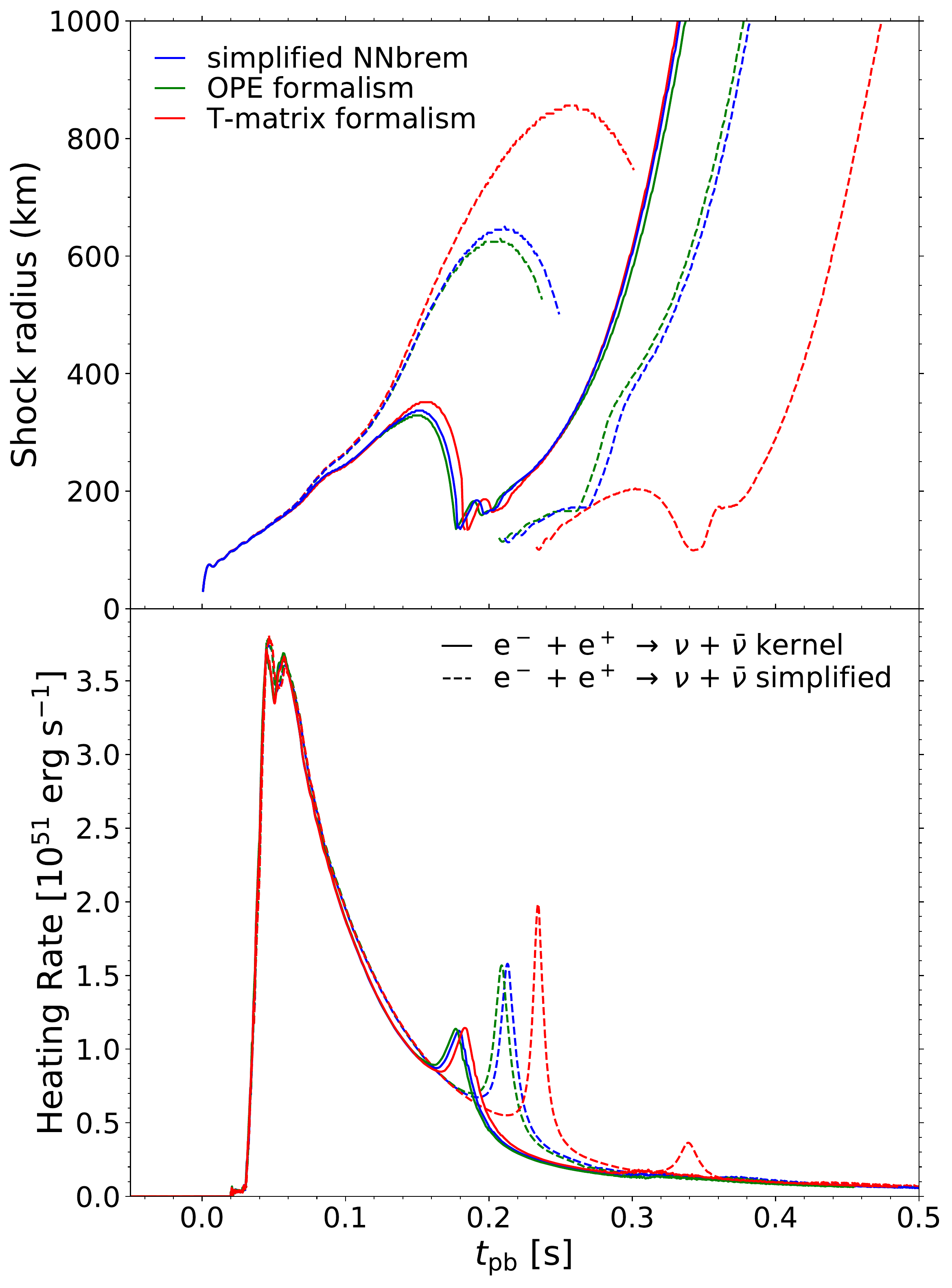}
\caption{The evolution of the shock radius (top panel) and neutrino heating (bottom panel) for the 9.6-$M_\odot$ progenitor vs. time. 
}
\label{2nd_shock_radius}
\end{figure}

\begin{figure}
\includegraphics[width=0.48\textwidth]{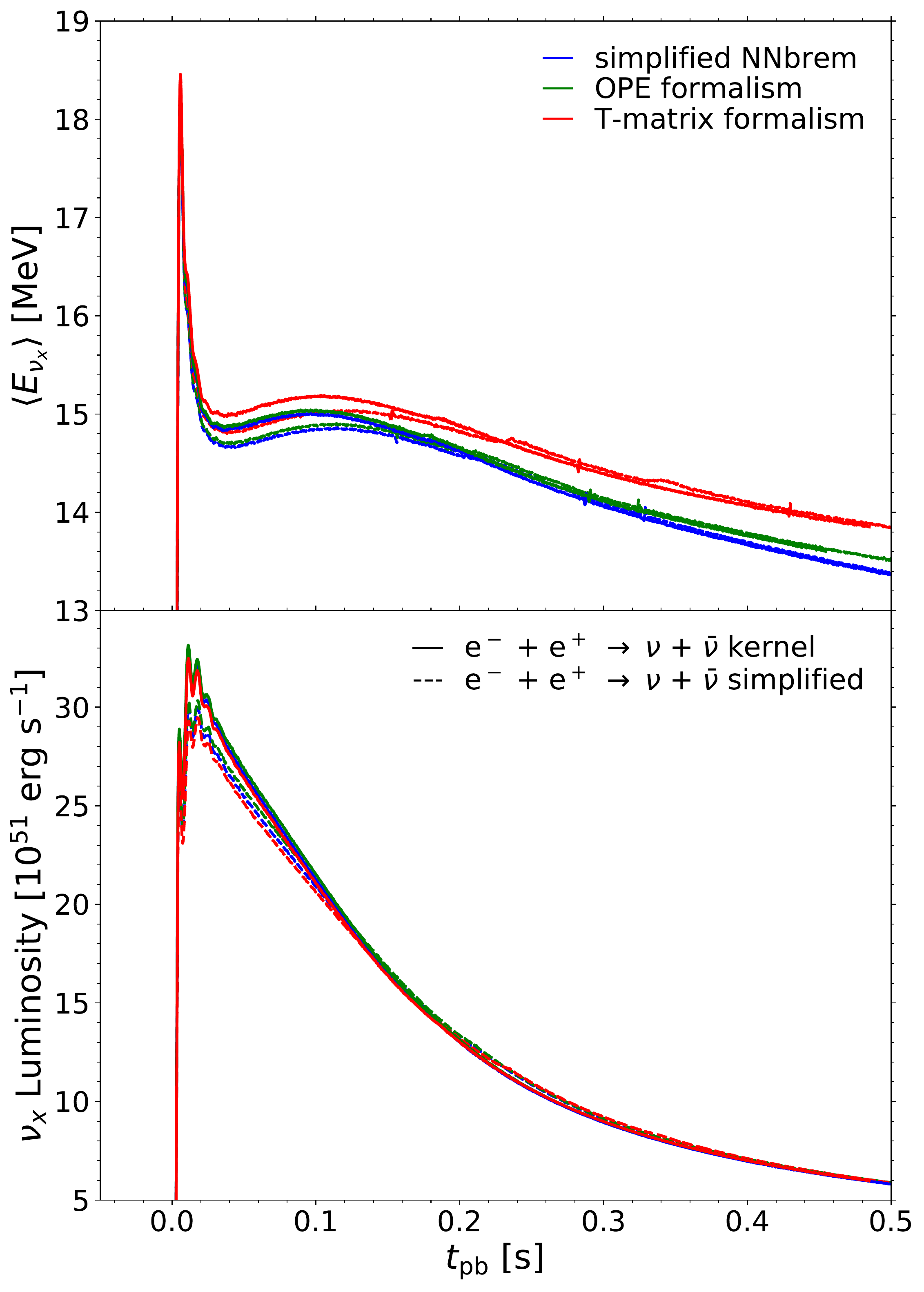}
\caption{Heavy-lepton neutrino quantities for the six models of the 9.6-M$_{\odot}$ progenitor with varying neutrino interactions. We show the mean energy and luminosity evolution in the top and bottom panels, respectively.}
\label{2nd_ave_energ}
\end{figure}

In order to show the impact of the different heavy-lepton pair process neutrino treatments on an exploding model, we evolved one of the few progenitors known to explode in 1D \cite{melson_neutrino-driven_2015, radice_electron-capture_2017}, a 9.6$M_\odot$, zero-metallicity star evolved with the KEPLER stellar evolution code \cite{heger:pc18}.   We simulate, using GR1D, the six combinations of thermal-interaction models listed in Table~\ref{run_list}.  All models successfully explode.   For all models, the shock expansion does not stall as in traditional iron-core collapse progenitors like it does in Fig.~\ref{s20_luminosities} for the 20M$_\odot$ progenitor. The shock radius evolution can be seen in the top panel of Fig.~\ref{2nd_shock_radius}. Here we plot the location of the maximum velocity gradient which flags accurately the position of the, sometimes multiple, different shocks.  The explosion times vary across the models, showing the sensitivity of the explosion to these neutrino interactions. The earliest explosion time, here arbitrarily defined as when the shock passes 1000\,km for the last time, is $\sim$310\,ms after bounce while the latest explosion is $\sim$470\,ms after bounce, $\sim$50\% longer. It is also notable that the initial shock formed at bounce is not the one that ultimately leads to the final explosion. In all cases, the first shock expands but is not energetic enough to runaway. The accretion of this material onto the PNS creates a burst of neutrino heating (bottom panel of Fig.~\ref{2nd_shock_radius}) and a secondary shock which will ultimately lead to an explosion. The formation time of this second shock corresponds to the heating peaks in the lower panel as the in-falling matter is compressed and the neutrino heating increases. In general, the models using the simplified treatment for electron-positron annihilation show a strong initial shock expansion phase, but a later ultimate explosion time, while the models with the full kernel treatment of electron-positron annihilation show a lower initial shock expansion and an earlier explosion.  We discuss the difference seen between the different interaction models in more detail below. 

\begin{figure*}
    \centering
    \includegraphics[width=\textwidth]{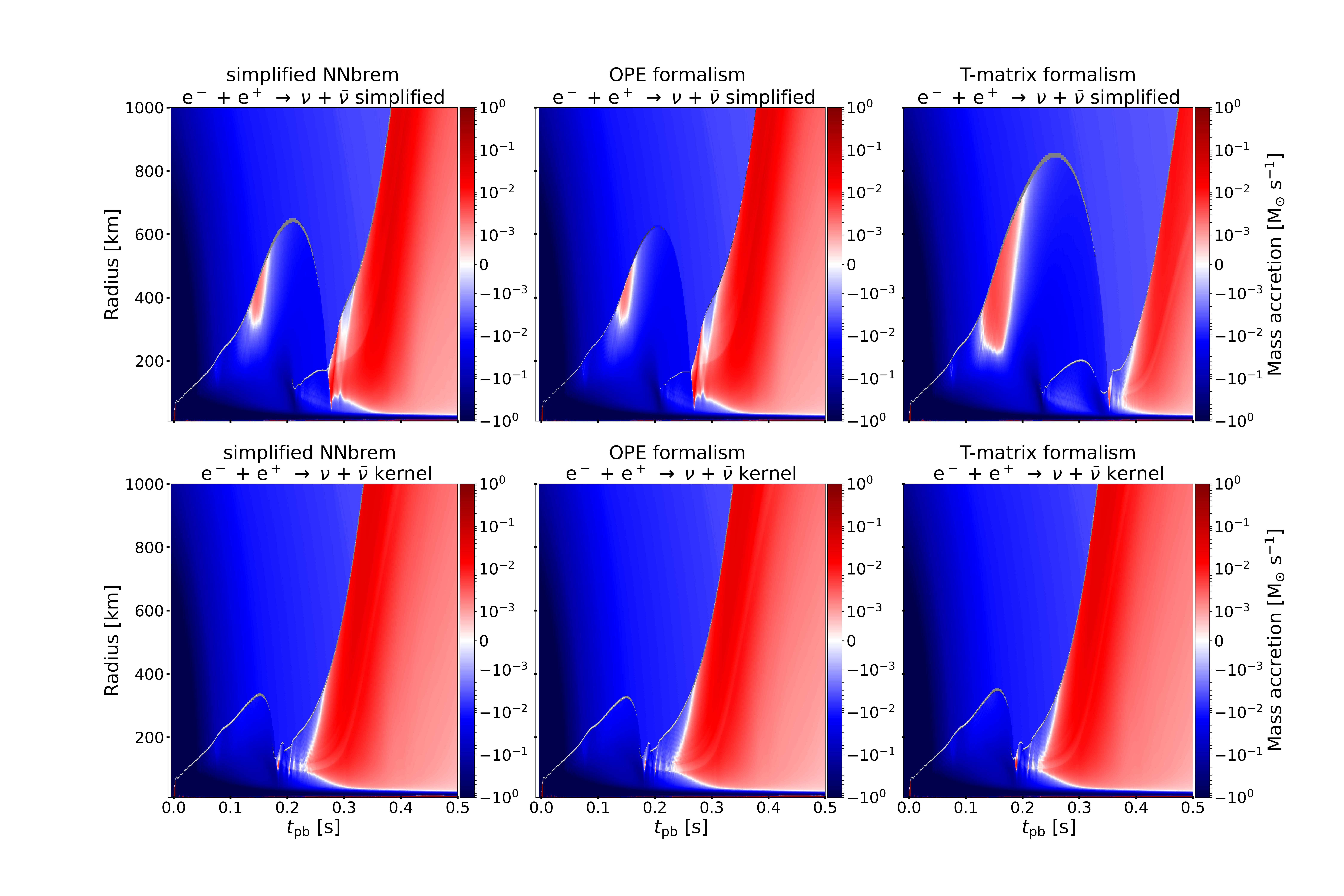}
    \caption{Mass accretion rates for the six different interaction sets explored vs. time. Blue colors denote accreting material, red denotes expanding material. Overlaid in grey are the location of the (sometimes multiple) shock fronts.}
\label{mass_acc}
\end{figure*}

In Fig.~\ref{2nd_ave_energ}, we show the heavy-lepton neutrino quantities for this progenitor model. In the top and bottom panel we show the mean energy and luminosity, respectively for each of the six neutrino pair-production treatments shown in Table.~\ref{run_list}. After a sharp and short peak at bounce, the mean energies plateau for the first $\sim$100\,ms after bounce around 15\,MeV with a spread $\sim$0.5\,MeV. The mean energies decrease over the remaining 400\,ms of the simulations by $\sim$1-1.5\,MeV, depending on the treatment used, with the T-matrix treatment maintaining the highest mean energy, as was the case in the 20M$_\odot$ progenitor above. The luminosities present a peak just after bounce and then slowly decrease until the end of the simulation. The heavy-lepton neutrino luminosities and energies present a similar dependence on the explored interactions as observed for the 20M$_{\odot}$ progenitor with the models using the full kernel treatment of the electron-positron pair annihilation having higher luminosities soon after bounce and then reaching similar, but slightly lower, values to the simplified formalism at later times.  As for the case of the 20-$M_\odot$ progenitor, we attribute these differences to the treatment of the transport in the simplified models. Particularly, the differences at early times are attributed to the form of the effective absorption coefficient which incorrectly treats the distribution of the annihilation partner and the smaller difference at late times due to the assumption of no final state blocking for the neutrinos during the emission process. At late times, the largest difference in the neutrino quantities arises from the use of the T-matrix bremsstrahlung kernels where the lower opacity at higher densities gives rise to higher neutrino energies, by $\sim$3\%, this is seen in both the simplified and the full kernel treatment.

We conclude our discussion of the 9.6-$M_\odot$ progenitor by examining the impact of the different neutrino interaction models in Tab.~\ref{run_list} on the development of the explosion and the shock propagation. The largest systematic difference we observe between the models using different interactions is the shock evolution for the three simulations that use the simplified treatment of electron-positron annihilation to neutrino pairs versus the three models that use the kernel-based treatment.  This is seen in Fig.~\ref{2nd_shock_radius}, but we show further evidence for this in Fig.~\ref{mass_acc}. In the six panels of Fig.~\ref{mass_acc}, we show the accretion history for each simulation.  Blue colors denote negative accretion rates (infalling material) while red colors show positive accretion rates (expanding material).  The three models for which we use the simplified treatment of electron-positron annihilation (models 1, 2, 3) are shown on the top row while the three models using the full kernel treatment (models 4, 5, 6) are shown on the bottom row. The left, middle, and right columns include the simplified bremsstrahlung treatment, the OPE kernel treatment, and the T-matrix kernel treatment, respectively. The three models with the simplified electron-positron annihilation pair production treatment have a slightly faster initial shock expansion from higher neutrino heating (see Fig.~\ref{2nd_shock_radius}), with the T-matrix bremsstrahlung treatment having the largest such expansion.  This causes these three models to undergo a strong initial shock acceleration phase starting around $\sim$130\,ms after bounce. Although, except for a small region directly behind the shock during this expansion phase, matter is mainly accreting in the postshock region.  These shocks eventually fail. It is worth noting that in multidimensional simulations of this progenitor \cite{melson_neutrino-driven_2015, radice_electron-capture_2017} the explosions typically set in during this period as the added role that multidimensional effects like convection and turbulence play is enough to initiate the explosion. However, in our spherically symmetric simulations this is not the case. Eventually, secondary shocks form at the surface of the PNS between $\sim$200-240\,ms after bounce concomitant with the increased accretion rate from the failing shock (see the dark blue regions around $\sim$100-150\,km at this time) which eventually give the ultimate explosion and the beginnings of a neutrino driven wind. The three models with the full kernel treatment for electron-positron annihilation do not undergo this accelerated expansion and continue to mildly expand until $\sim$160\,ms at which point the shock fails and the secondary shock forms at $\sim$180\,ms after bounce. These models are the first to ultimately explode. 

\begin{figure}
    \centering
     \includegraphics[width=0.5\textwidth]{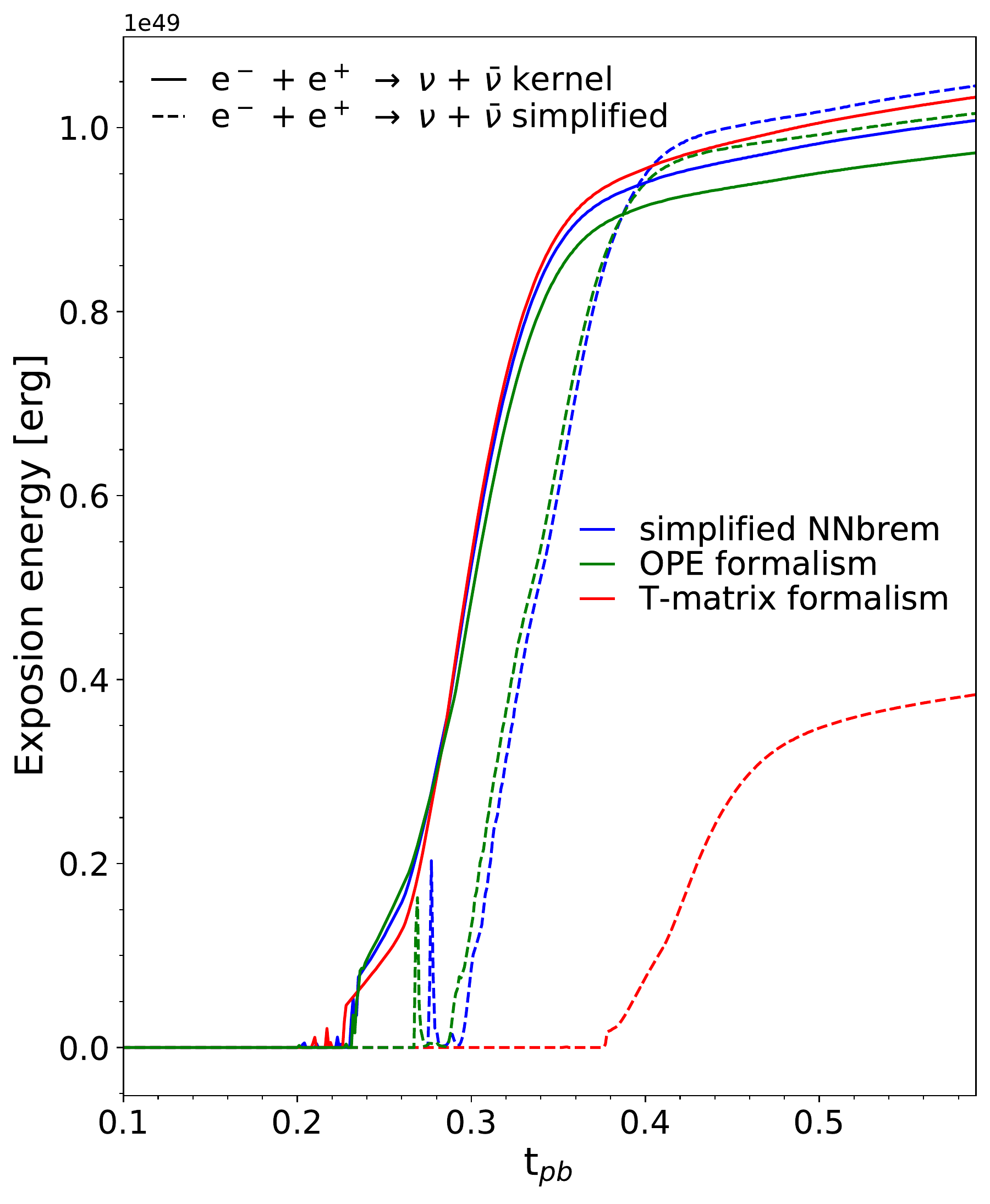}
    \caption{Diagnostic explosion energy for the 9.6-$M_\odot$ progenitor model for each of the neutrino interaction sets explored.  Early explosions generally give higher diagnostic explosion energies.}
    \label{explo_energ}
\end{figure}

We compute the diagnostic explosion energy for each of the six models simulated,
\begin{equation}
    E_\mathrm{dia} = \int_{v>0} \left[\phi + \frac{v^2}{2} + (\epsilon-\epsilon_0) \right]_{>0} dm\,,\label{eq:diaE}
\end{equation}
where $\phi$ is the gravitational potential, $v$ is the fluid velocity, $\epsilon$ is the specific internal energy, and $\epsilon_0$ is a reference zero-point taken for simplicity to be the value of the internal energy of the EOS for the same density and $Y_e$ but for $T=0$. We only consider contributions to the diagnostic energy from outflowing matter and where the integrand of Eq.~\ref{eq:diaE} is positive, this is a proxy for unbound material.  The results are shown Fig.\ref{explo_energ}. In general, for the 9.6-$M_\odot$ progenitor the explosion energy depends on the time of the explosion.  For early explosions in multiple dimensions, for example see \cite{melson_neutrino-driven_2015, radice_electron-capture_2017}, the explosion energies can reach several to 10 times $10^{49}$\,erg. In our relatively late spherically-symmetric explosions, we see explosion energies $\sim10^{49}$\,erg for the five models that explode first. The last model to explode achieves only $0.4\times10^{49}$\,erg owing to the much lower neutrino heating rates present at the later times. We see that while the rise of the explosion energy is strongly correlated with the onset of the explosion, the ultimate explosion energy is not strictly dependent on the explosion time.  We see that differences can arise due to the complex interplay of the neutrino heating of the fallback material from the initial failed shock (see Fig.~\ref{mass_acc}).

Finally, we note that $\epsilon - \epsilon_0$ in Eq.~\ref{eq:diaE} is an estimate of the sum of the thermal energies and the recombination energy that will be converted to thermal energy as the matter recombines.  We tested this against an explicit calculation similar to that of \cite{melson_neutrino-driven_2015}.  If we assume that all the free nucleons and alpha particles present in the matter would recombine to iron this sets the specific recombination energy.  We use this, along with the specific internal energy from the Helmholtz EOS \cite{timmes:99} at a matter temperature of $T$ to replace $\epsilon - \epsilon_0$ in Eq.~\ref{eq:diaE}.  We find comparable, to within 5\%, diagnostic explosion energies as shown in Fig.~\ref{explo_energ}.

\section{\label{sec:conclusion}Conclusion}

In this paper, we highlighted and explored the importance of heavy-lepton ($\nu_\mu$, $\nu_\tau$, and their antiparticles) neutrinos and the interactions that produce them in the scope of spherically-symmetry, fully general-relativistic neutrino-driven CCSNe simulations. We performed systematic simulations on two different progenitors, a solar metallicity progenitor star with a ZAMS mass of 20M$_{\odot}$ which has been extensively used in the literature for CCSN simulations, including a recent comparison study and a zero metallicity progenitor star with a ZAMS mass of 9.6M$_{\odot}$ that has the peculiarity to explode even in spherically symmetric simulations.  We simulate the core collapse and the early post-bounce phase using the open-source software GR1D and Nulib, which we update accordingly with the work presented in this paper. In particular, we test the importance of the two main heavy-lepton neutrino pair-production processes in CCSNe, electron-positron annihilation to a neutrino-antineutrino pair and nucleon-nucleon scattering that radiates a neutrino-antineutrino pair (i.e. nucleon-nucleon bremsstrahlung).  We explore two main effects.  First, we study the neutrino transport implementation of both of these interactions by utilizing a simplified approached with effective emission and absorption coefficients and a complete treatment utilizing complete scattering kernels and in situ neutrino distribution functions. The aim for this part of the study is to assess the impact and quantify systematic effects of the simplified treatment on CCSN simulations with the goal of providing a robust prescription for use in multidimensional simulations.  Second, we explored two independent nuclear-physics based prescriptions for the nucleon-nucleon bremsstrahlung interaction (in addition to a simplified approach as mentioned above). One of these interactions is commonly used throughout the literature for CCSNe simulations and is based on the one-pion exchange formalism \cite{hannestad_supernova_1998}.  The other interaction \cite{guo_chiral_2019}, formulates the nucleon-nucleon bremsstrahlung scattering kernel based on a T-matrix formalism where the underlying interaction is constrained by experimental phase shifts, see also \cite{bartl_supernova_2014}. From these variations we arrive at six combinations of interactions to explore with our CCSN simulations.

We find that overall the simplified neutrino interactions do a fair job at reproducing the neutrino quantities (within $\sim$10\%) and the dynamics of the core collapse event, although potentially important differences are present that need to be considered when employing the simplified treatment for precision simulations.  We find the simplified method under predicts the heavy-lepton neutrino luminosity in the early stages, by $\sim$10\%.  We show that this is the result of the simplistic treatment of the distribution of the annihilation partner, i.e. the assumption that it follows the blackbody distribution.  This leads to an overestimate of the annihilation rate as the neutrinos begin to free stream away from the CCSN core. In the region where this excess annihilation is occurring the distribution function is quite isotropic and therefore, as opposed to that suggested in previous works, this difference is unlikely due to the assumption of isotropy in the simplified treatment.  While the largest impact is seen with different treatments for the electron-positron annihilation interaction, we see the same (but much smaller in impact) trend with the simplified treatment of nucleon-nucleon bremsstrahlung. At later times the simplified treatment agrees much better with the full kernel treatment, owing to the fact that the distribution of the annihilation partner is much closer to the blackbody distribution in the regions where the neutrinos are decoupling. We do see a slight overestimate of the luminosity for the simplified treatment, which we attribute to the assumption of no final-state neutrino blocking in the simplified emission treatment. Finally, we comment on the impact of using a different microphysical interaction for the nucleon-nucleon bremsstrahlung. The use of the T-matrix formalism over the standard one-pion exchange treatment systematically increases the heavy-lepton neutrino energy by $\sim$5\% which we infer is due to the reduced interaction strength of the T-matrix kernel at larger densities compared the one-pion exchange kernel.  This causes the heavy-lepton neutrinos to begin to decouple deeper into the PNS core, where the temperature is higher.  

For cases where the dynamics can sensitively depend on the neutrino physics, for example with the 9.6M$_\odot$ progenitor studied here, we find that the different interactions explored can impact the heating enough at the earlier stages to quantitatively effect the development of the explosion, including the ultimate explosion time and the explosion energy.  To what extent this carries over to multidimensional simulations remains to be seen, although it is important to note that even in multidimensional simulations we know there is sensitivity to the neutrino physics at the $\sim$10\% level demonstrated here \cite{melson_neutrino-driven_2015-1, oconnor_core-collapse_2017}.

\section*{\label{sec:Acknowledgment}Acknowledgment}

We wish to thanks Gang Guo for supplying us with tables for T-matrix treatment critical for the work presented here. We also wish to thank Sean Couch, Andre da Silva Schneider, Kei Kotake, Sanjay Reddy, Achim Schwenk, and Shuai Zha for the fruitful discussions. The authors would like to acknowledge Vetenskapsr{\aa}det (the Swedish Research Council) for supporting this work under award number 2018-04575. The computations were enabled by resources provided by the Swedish National Infrastructure for Computing (SNIC) at NSC and PDC partially funded by the Swedish Research Council through grant agreement no. 2018-0597. We also acknowledge support from the “ChETEC” COST Action (CA16117), supported by COST (European Cooperation in Science and Technology)

\bibliography{total_library.bib}

\begin{thebibliography}{57}%
\makeatletter
\providecommand \@ifxundefined [1]{%
 \@ifx{#1\undefined}
}%
\providecommand \@ifnum [1]{%
 \ifnum #1\expandafter \@firstoftwo
 \else \expandafter \@secondoftwo
 \fi
}%
\providecommand \@ifx [1]{%
 \ifx #1\expandafter \@firstoftwo
 \else \expandafter \@secondoftwo
 \fi
}%
\providecommand \natexlab [1]{#1}%
\providecommand \enquote  [1]{``#1''}%
\providecommand \bibnamefont  [1]{#1}%
\providecommand \bibfnamefont [1]{#1}%
\providecommand \citenamefont [1]{#1}%
\providecommand \href@noop [0]{\@secondoftwo}%
\providecommand \href [0]{\begingroup \@sanitize@url \@href}%
\providecommand \@href[1]{\@@startlink{#1}\@@href}%
\providecommand \@@href[1]{\endgroup#1\@@endlink}%
\providecommand \@sanitize@url [0]{\catcode `\\12\catcode `\$12\catcode
  `\&12\catcode `\#12\catcode `\^12\catcode `\_12\catcode `\%12\relax}%
\providecommand \@@startlink[1]{}%
\providecommand \@@endlink[0]{}%
\providecommand \url  [0]{\begingroup\@sanitize@url \@url }%
\providecommand \@url [1]{\endgroup\@href {#1}{\urlprefix }}%
\providecommand \urlprefix  [0]{URL }%
\providecommand \Eprint [0]{\href }%
\providecommand \doibase [0]{http://dx.doi.org/}%
\providecommand \selectlanguage [0]{\@gobble}%
\providecommand \bibinfo  [0]{\@secondoftwo}%
\providecommand \bibfield  [0]{\@secondoftwo}%
\providecommand \translation [1]{[#1]}%
\providecommand \BibitemOpen [0]{}%
\providecommand \bibitemStop [0]{}%
\providecommand \bibitemNoStop [0]{.\EOS\space}%
\providecommand \EOS [0]{\spacefactor3000\relax}%
\providecommand \BibitemShut  [1]{\csname bibitem#1\endcsname}%
\let\auto@bib@innerbib\@empty
%</preamble>
\bibitem [{\citenamefont {Woosley}\ and\ \citenamefont
  {Heger}(2002)}]{woosley_evolution_2002}%
  \BibitemOpen
  \bibfield  {author} {\bibinfo {author} {\bibfnamefont {S.}~\bibnamefont
  {Woosley}}\ and\ \bibinfo {author} {\bibfnamefont {A.}~\bibnamefont
  {Heger}},\ }\href
  {https://journals.aps.org/rmp/abstract/10.1103/RevModPhys.74.1015} {\bibfield
   {journal} {\bibinfo  {journal} {Rev. Mod. Phys.}\ }\textbf {\bibinfo
  {volume} {74}} (\bibinfo {year} {2002})}\BibitemShut {NoStop}%
\bibitem [{\citenamefont {Thielemann}\ \emph {et~al.}(2017)\citenamefont
  {Thielemann}, \citenamefont {Eichler}, \citenamefont {Panov},\ and\
  \citenamefont {Wehmeyer}}]{thielemann_neutron_2017}%
  \BibitemOpen
  \bibfield  {author} {\bibinfo {author} {\bibfnamefont {F.-K.}\ \bibnamefont
  {Thielemann}}, \bibinfo {author} {\bibfnamefont {M.}~\bibnamefont {Eichler}},
  \bibinfo {author} {\bibfnamefont {I.~V.}\ \bibnamefont {Panov}}, \ and\
  \bibinfo {author} {\bibfnamefont {B.}~\bibnamefont {Wehmeyer}},\ }\href
  {\doibase 10.1146/annurev-nucl-101916-123246} {\bibfield  {journal} {\bibinfo
   {journal} {Annu. Rev. Nucl. Part. Sci.}\ }\textbf {\bibinfo {volume} {67}},\
  \bibinfo {pages} {253} (\bibinfo {year} {2017})},\ \bibinfo {note} {arXiv:
  1710.02142}\BibitemShut {NoStop}%
\bibitem [{\citenamefont {Bruch}\ \emph {et~al.}(2020)\citenamefont {Bruch},
  \citenamefont {Gal-Yam}, \citenamefont {Schulze}, \citenamefont {Yaron},
  \citenamefont {Yang}, \citenamefont {Soumagnac}, \citenamefont {Rigault},
  \citenamefont {Strotjohann}, \citenamefont {Ofek}, \citenamefont {Sollerman},
  \citenamefont {Masci}, \citenamefont {Barbarino}, \citenamefont {Ho},
  \citenamefont {Fremling}, \citenamefont {Perley}, \citenamefont {Nordin},
  \citenamefont {Cenko}, \citenamefont {Adams}, \citenamefont {Adreoni},
  \citenamefont {Bellm}, \citenamefont {Blagorodnova}, \citenamefont {Bulla},
  \citenamefont {Burdge}, \citenamefont {De}, \citenamefont {Dhawan},
  \citenamefont {Drake}, \citenamefont {Duev}, \citenamefont {Dugas},
  \citenamefont {Graham}, \citenamefont {Graham}, \citenamefont {Jencson},
  \citenamefont {Karamehmetoglu}, \citenamefont {Kasliwal}, \citenamefont
  {Kim}, \citenamefont {Kulkarni}, \citenamefont {Kupfer}, \citenamefont
  {Mahabal}, \citenamefont {Miller}, \citenamefont {Prince}, \citenamefont
  {Riddle}, \citenamefont {Sharma}, \citenamefont {Smith}, \citenamefont
  {Taddia}, \citenamefont {Taggart}, \citenamefont {Walters},\ and\
  \citenamefont {Yan}}]{bruch_large_2020}%
  \BibitemOpen
  \bibfield  {author} {\bibinfo {author} {\bibfnamefont {R.~J.}\ \bibnamefont
  {Bruch}}, \bibinfo {author} {\bibfnamefont {A.}~\bibnamefont {Gal-Yam}},
  \bibinfo {author} {\bibfnamefont {S.}~\bibnamefont {Schulze}}, \bibinfo
  {author} {\bibfnamefont {O.}~\bibnamefont {Yaron}}, \bibinfo {author}
  {\bibfnamefont {Y.}~\bibnamefont {Yang}}, \bibinfo {author} {\bibfnamefont
  {M.~T.}\ \bibnamefont {Soumagnac}}, \bibinfo {author} {\bibfnamefont
  {M.}~\bibnamefont {Rigault}}, \bibinfo {author} {\bibfnamefont {N.~L.}\
  \bibnamefont {Strotjohann}}, \bibinfo {author} {\bibfnamefont
  {E.}~\bibnamefont {Ofek}}, \bibinfo {author} {\bibfnamefont {J.}~\bibnamefont
  {Sollerman}}, \bibinfo {author} {\bibfnamefont {F.~J.}\ \bibnamefont
  {Masci}}, \bibinfo {author} {\bibfnamefont {C.}~\bibnamefont {Barbarino}},
  \bibinfo {author} {\bibfnamefont {A.~Y.~Q.}\ \bibnamefont {Ho}}, \bibinfo
  {author} {\bibfnamefont {C.}~\bibnamefont {Fremling}}, \bibinfo {author}
  {\bibfnamefont {D.}~\bibnamefont {Perley}}, \bibinfo {author} {\bibfnamefont
  {J.}~\bibnamefont {Nordin}}, \bibinfo {author} {\bibfnamefont {S.~B.}\
  \bibnamefont {Cenko}}, \bibinfo {author} {\bibfnamefont {S.}~\bibnamefont
  {Adams}}, \bibinfo {author} {\bibfnamefont {I.}~\bibnamefont {Adreoni}},
  \bibinfo {author} {\bibfnamefont {E.~C.}\ \bibnamefont {Bellm}}, \bibinfo
  {author} {\bibfnamefont {N.}~\bibnamefont {Blagorodnova}}, \bibinfo {author}
  {\bibfnamefont {M.}~\bibnamefont {Bulla}}, \bibinfo {author} {\bibfnamefont
  {K.}~\bibnamefont {Burdge}}, \bibinfo {author} {\bibfnamefont
  {K.}~\bibnamefont {De}}, \bibinfo {author} {\bibfnamefont {S.}~\bibnamefont
  {Dhawan}}, \bibinfo {author} {\bibfnamefont {A.~J.}\ \bibnamefont {Drake}},
  \bibinfo {author} {\bibfnamefont {D.~A.}\ \bibnamefont {Duev}}, \bibinfo
  {author} {\bibfnamefont {A.}~\bibnamefont {Dugas}}, \bibinfo {author}
  {\bibfnamefont {M.}~\bibnamefont {Graham}}, \bibinfo {author} {\bibfnamefont
  {M.~L.}\ \bibnamefont {Graham}}, \bibinfo {author} {\bibfnamefont
  {J.}~\bibnamefont {Jencson}}, \bibinfo {author} {\bibfnamefont
  {E.}~\bibnamefont {Karamehmetoglu}}, \bibinfo {author} {\bibfnamefont
  {M.}~\bibnamefont {Kasliwal}}, \bibinfo {author} {\bibfnamefont {Y.-L.}\
  \bibnamefont {Kim}}, \bibinfo {author} {\bibfnamefont {S.}~\bibnamefont
  {Kulkarni}}, \bibinfo {author} {\bibfnamefont {T.}~\bibnamefont {Kupfer}},
  \bibinfo {author} {\bibfnamefont {A.}~\bibnamefont {Mahabal}}, \bibinfo
  {author} {\bibfnamefont {A.~A.}\ \bibnamefont {Miller}}, \bibinfo {author}
  {\bibfnamefont {T.~A.}\ \bibnamefont {Prince}}, \bibinfo {author}
  {\bibfnamefont {R.}~\bibnamefont {Riddle}}, \bibinfo {author} {\bibfnamefont
  {Y.}~\bibnamefont {Sharma}}, \bibinfo {author} {\bibfnamefont
  {R.}~\bibnamefont {Smith}}, \bibinfo {author} {\bibfnamefont
  {F.}~\bibnamefont {Taddia}}, \bibinfo {author} {\bibfnamefont
  {K.}~\bibnamefont {Taggart}}, \bibinfo {author} {\bibfnamefont
  {R.}~\bibnamefont {Walters}}, \ and\ \bibinfo {author} {\bibfnamefont
  {L.}~\bibnamefont {Yan}},\ }\href {http://arxiv.org/abs/2008.09986}
  {\bibfield  {journal} {\bibinfo  {journal} {arXiv:2008.09986 [astro-ph]}\ }
  (\bibinfo {year} {2020})},\ \bibinfo {note} {arXiv: 2008.09986}\BibitemShut
  {NoStop}%
\bibitem [{\citenamefont {Janka}(2017{\natexlab{a}})}]{janka_neutrino_2017}%
  \BibitemOpen
  \bibfield  {author} {\bibinfo {author} {\bibfnamefont {H.-T.}\ \bibnamefont
  {Janka}},\ }in\ \href {\doibase 10.1007/978-3-319-21846-5_4} {\emph {\bibinfo
  {booktitle} {Handbook of Supernovae}}},\ \bibinfo {editor} {edited by\
  \bibinfo {editor} {\bibfnamefont {A.~W.}\ \bibnamefont {Alsabti}}\ and\
  \bibinfo {editor} {\bibfnamefont {P.}~\bibnamefont {Murdin}}}\ (\bibinfo
  {publisher} {Springer International Publishing},\ \bibinfo {address} {Cham},\
  \bibinfo {year} {2017})\ pp.\ \bibinfo {pages} {1575--1604}\BibitemShut
  {NoStop}%
\bibitem [{\citenamefont {Kotake}\ and\ \citenamefont
  {Kuroda}(2017)}]{kotake_gravitational_2017}%
  \BibitemOpen
  \bibfield  {author} {\bibinfo {author} {\bibfnamefont {K.}~\bibnamefont
  {Kotake}}\ and\ \bibinfo {author} {\bibfnamefont {T.}~\bibnamefont
  {Kuroda}},\ }in\ \href {\doibase 10.1007/978-3-319-21846-5_9} {\emph
  {\bibinfo {booktitle} {Handbook of Supernovae}}},\ \bibinfo {editor} {edited
  by\ \bibinfo {editor} {\bibfnamefont {A.~W.}\ \bibnamefont {Alsabti}}\ and\
  \bibinfo {editor} {\bibfnamefont {P.}~\bibnamefont {Murdin}}}\ (\bibinfo
  {publisher} {Springer International Publishing},\ \bibinfo {address} {Cham},\
  \bibinfo {year} {2017})\ pp.\ \bibinfo {pages} {1671--1698}\BibitemShut
  {NoStop}%
\bibitem [{\citenamefont {{O'Connor}}\ \emph {et~al.}(2018)\citenamefont
  {{O'Connor}}, \citenamefont {{Bollig}}, \citenamefont {{Burrows}},
  \citenamefont {{Couch}}, \citenamefont {{Fischer}}, \citenamefont {{Janka}},
  \citenamefont {{Kotake}}, \citenamefont {{Lentz}}, \citenamefont
  {{Liebend{\"o}rfer}}, \citenamefont {{Messer}}, \citenamefont {{Mezzacappa}},
  \citenamefont {{Takiwaki}},\ and\ \citenamefont
  {{Vartanyan}}}]{oconnor_global_2018}%
  \BibitemOpen
  \bibfield  {author} {\bibinfo {author} {\bibfnamefont {E.}~\bibnamefont
  {{O'Connor}}}, \bibinfo {author} {\bibfnamefont {R.}~\bibnamefont
  {{Bollig}}}, \bibinfo {author} {\bibfnamefont {A.}~\bibnamefont {{Burrows}}},
  \bibinfo {author} {\bibfnamefont {S.}~\bibnamefont {{Couch}}}, \bibinfo
  {author} {\bibfnamefont {T.}~\bibnamefont {{Fischer}}}, \bibinfo {author}
  {\bibfnamefont {H.-T.}\ \bibnamefont {{Janka}}}, \bibinfo {author}
  {\bibfnamefont {K.}~\bibnamefont {{Kotake}}}, \bibinfo {author}
  {\bibfnamefont {E.~J.}\ \bibnamefont {{Lentz}}}, \bibinfo {author}
  {\bibfnamefont {M.}~\bibnamefont {{Liebend{\"o}rfer}}}, \bibinfo {author}
  {\bibfnamefont {O.~E.~B.}\ \bibnamefont {{Messer}}}, \bibinfo {author}
  {\bibfnamefont {A.}~\bibnamefont {{Mezzacappa}}}, \bibinfo {author}
  {\bibfnamefont {T.}~\bibnamefont {{Takiwaki}}}, \ and\ \bibinfo {author}
  {\bibfnamefont {D.}~\bibnamefont {{Vartanyan}}},\ }\href {\doibase
  10.1088/1361-6471/aadeae} {\bibfield  {journal} {\bibinfo  {journal} {Journal
  of Physics G Nuclear Physics}\ }\textbf {\bibinfo {volume} {45}},\ \bibinfo
  {pages} {104001} (\bibinfo {year} {2018})},\ \Eprint
  {http://arxiv.org/abs/1806.04175} {arXiv:1806.04175 [astro-ph.HE]}
  \BibitemShut {NoStop}%
\bibitem [{\citenamefont {Mösta}\ \emph {et~al.}(2014)\citenamefont {Mösta},
  \citenamefont {Richers}, \citenamefont {Ott}, \citenamefont {Haas},
  \citenamefont {Piro}, \citenamefont {Boydstun}, \citenamefont {Abdikamalov},
  \citenamefont {Reisswig},\ and\ \citenamefont
  {Schnetter}}]{mosta_magnetorotational_2014}%
  \BibitemOpen
  \bibfield  {author} {\bibinfo {author} {\bibfnamefont {P.}~\bibnamefont
  {Mösta}}, \bibinfo {author} {\bibfnamefont {S.}~\bibnamefont {Richers}},
  \bibinfo {author} {\bibfnamefont {C.~D.}\ \bibnamefont {Ott}}, \bibinfo
  {author} {\bibfnamefont {R.}~\bibnamefont {Haas}}, \bibinfo {author}
  {\bibfnamefont {A.~L.}\ \bibnamefont {Piro}}, \bibinfo {author}
  {\bibfnamefont {K.}~\bibnamefont {Boydstun}}, \bibinfo {author}
  {\bibfnamefont {E.}~\bibnamefont {Abdikamalov}}, \bibinfo {author}
  {\bibfnamefont {C.}~\bibnamefont {Reisswig}}, \ and\ \bibinfo {author}
  {\bibfnamefont {E.}~\bibnamefont {Schnetter}},\ }\href {\doibase
  10.1088/2041-8205/785/2/L29} {\bibfield  {journal} {\bibinfo  {journal}
  {ApJ}\ }\textbf {\bibinfo {volume} {785}},\ \bibinfo {pages} {L29} (\bibinfo
  {year} {2014})},\ \bibinfo {note} {arXiv: 1403.1230}\BibitemShut {NoStop}%
\bibitem [{\citenamefont {{Glas}}\ \emph {et~al.}(2020)\citenamefont {{Glas}},
  \citenamefont {{Janka}}, \citenamefont {{Capozzi}}, \citenamefont {{Sen}},
  \citenamefont {{Dasgupta}}, \citenamefont {{Mirizzi}},\ and\ \citenamefont
  {{Sigl}}}]{glas_fast_2019}%
  \BibitemOpen
  \bibfield  {author} {\bibinfo {author} {\bibfnamefont {R.}~\bibnamefont
  {{Glas}}}, \bibinfo {author} {\bibfnamefont {H.~T.}\ \bibnamefont {{Janka}}},
  \bibinfo {author} {\bibfnamefont {F.}~\bibnamefont {{Capozzi}}}, \bibinfo
  {author} {\bibfnamefont {M.}~\bibnamefont {{Sen}}}, \bibinfo {author}
  {\bibfnamefont {B.}~\bibnamefont {{Dasgupta}}}, \bibinfo {author}
  {\bibfnamefont {A.}~\bibnamefont {{Mirizzi}}}, \ and\ \bibinfo {author}
  {\bibfnamefont {G.}~\bibnamefont {{Sigl}}},\ }\href {\doibase
  10.1103/PhysRevD.101.063001} {\bibfield  {journal} {\bibinfo  {journal}
  {\prd}\ }\textbf {\bibinfo {volume} {101}},\ \bibinfo {eid} {063001}
  (\bibinfo {year} {2020})},\ \Eprint {http://arxiv.org/abs/1912.00274}
  {arXiv:1912.00274 [astro-ph.HE]} \BibitemShut {NoStop}%
\bibitem [{\citenamefont {Burrows}\ \emph {et~al.}(2019)\citenamefont
  {Burrows}, \citenamefont {Radice},\ and\ \citenamefont
  {Vartanyan}}]{burrows_three-dimensional_2019}%
  \BibitemOpen
  \bibfield  {author} {\bibinfo {author} {\bibfnamefont {A.}~\bibnamefont
  {Burrows}}, \bibinfo {author} {\bibfnamefont {D.}~\bibnamefont {Radice}}, \
  and\ \bibinfo {author} {\bibfnamefont {D.}~\bibnamefont {Vartanyan}},\ }\href
  {http://arxiv.org/abs/1902.00547} {\bibfield  {journal} {\bibinfo  {journal}
  {arXiv:1902.00547 [astro-ph]}\ } (\bibinfo {year} {2019})},\ \bibinfo {note}
  {arXiv: 1902.00547}\BibitemShut {NoStop}%
\bibitem [{\citenamefont {Vartanyan}\ \emph
  {et~al.}(2018{\natexlab{a}})\citenamefont {Vartanyan}, \citenamefont
  {Burrows}, \citenamefont {Radice}, \citenamefont {Skinner},\ and\
  \citenamefont {Dolence}}]{vartanyan_revival_2018}%
  \BibitemOpen
  \bibfield  {author} {\bibinfo {author} {\bibfnamefont {D.}~\bibnamefont
  {Vartanyan}}, \bibinfo {author} {\bibfnamefont {A.}~\bibnamefont {Burrows}},
  \bibinfo {author} {\bibfnamefont {D.}~\bibnamefont {Radice}}, \bibinfo
  {author} {\bibfnamefont {M.~A.}\ \bibnamefont {Skinner}}, \ and\ \bibinfo
  {author} {\bibfnamefont {J.}~\bibnamefont {Dolence}},\ }\href {\doibase
  10.1093/mnras/sty809} {\bibfield  {journal} {\bibinfo  {journal} {Monthly
  Notices of the Royal Astronomical Society}\ }\textbf {\bibinfo {volume}
  {477}},\ \bibinfo {pages} {3091} (\bibinfo {year} {2018}{\natexlab{a}})},\
  \bibinfo {note} {arXiv: 1801.08148}\BibitemShut {NoStop}%
\bibitem [{\citenamefont {Pan}\ \emph {et~al.}(2018)\citenamefont {Pan},
  \citenamefont {Mattes}, \citenamefont {O'Connor}, \citenamefont {Couch},
  \citenamefont {Perego},\ and\ \citenamefont {Arcones}}]{pan_impact_2018}%
  \BibitemOpen
  \bibfield  {author} {\bibinfo {author} {\bibfnamefont {K.-C.}\ \bibnamefont
  {Pan}}, \bibinfo {author} {\bibfnamefont {C.}~\bibnamefont {Mattes}},
  \bibinfo {author} {\bibfnamefont {E.~P.}\ \bibnamefont {O'Connor}}, \bibinfo
  {author} {\bibfnamefont {S.~M.}\ \bibnamefont {Couch}}, \bibinfo {author}
  {\bibfnamefont {A.}~\bibnamefont {Perego}}, \ and\ \bibinfo {author}
  {\bibfnamefont {A.}~\bibnamefont {Arcones}},\ }\href
  {http://arxiv.org/abs/1806.10030} {\bibfield  {journal} {\bibinfo  {journal}
  {arXiv:1806.10030 [astro-ph]}\ } (\bibinfo {year} {2018})},\ \bibinfo {note}
  {arXiv: 1806.10030}\BibitemShut {NoStop}%
\bibitem [{\citenamefont
  {Janka}(2017{\natexlab{b}})}]{janka_neutrino-driven_2017}%
  \BibitemOpen
  \bibfield  {author} {\bibinfo {author} {\bibfnamefont {H.-T.}\ \bibnamefont
  {Janka}},\ }in\ \href {\doibase 10.1007/978-3-319-21846-5_109} {\emph
  {\bibinfo {booktitle} {Handbook of Supernovae}}},\ \bibinfo {editor} {edited
  by\ \bibinfo {editor} {\bibfnamefont {A.~W.}\ \bibnamefont {Alsabti}}\ and\
  \bibinfo {editor} {\bibfnamefont {P.}~\bibnamefont {Murdin}}}\ (\bibinfo
  {publisher} {Springer International Publishing},\ \bibinfo {address} {Cham},\
  \bibinfo {year} {2017})\ pp.\ \bibinfo {pages} {1095--1150}\BibitemShut
  {NoStop}%
\bibitem [{\citenamefont {Bethe}\ and\ \citenamefont
  {Wilson}(1985)}]{bethe_revival_1985}%
  \BibitemOpen
  \bibfield  {author} {\bibinfo {author} {\bibfnamefont {H.~A.}\ \bibnamefont
  {Bethe}}\ and\ \bibinfo {author} {\bibfnamefont {J.~R.}\ \bibnamefont
  {Wilson}},\ }\href {\doibase 10.1086/163343} {\bibfield  {journal} {\bibinfo
  {journal} {The Astrophysical Journal}\ }\textbf {\bibinfo {volume} {295}},\
  \bibinfo {pages} {14} (\bibinfo {year} {1985})}\BibitemShut {NoStop}%
\bibitem [{\citenamefont {Bethe}(1990)}]{bethe_supernova_1990}%
  \BibitemOpen
  \bibfield  {author} {\bibinfo {author} {\bibfnamefont {H.~A.}\ \bibnamefont
  {Bethe}},\ }\href {\doibase 10.1103/RevModPhys.62.801} {\bibfield  {journal}
  {\bibinfo  {journal} {Rev. Mod. Phys.}\ }\textbf {\bibinfo {volume} {62}},\
  \bibinfo {pages} {801} (\bibinfo {year} {1990})},\ \bibinfo {note}
  {publisher: American Physical Society}\BibitemShut {NoStop}%
\bibitem [{\citenamefont {Janka}\ \emph {et~al.}(2012)\citenamefont {Janka},
  \citenamefont {Hanke}, \citenamefont {Huedepohl}, \citenamefont {Marek},
  \citenamefont {Mueller},\ and\ \citenamefont
  {Obergaulinger}}]{janka_core-collapse_2012}%
  \BibitemOpen
  \bibfield  {author} {\bibinfo {author} {\bibfnamefont {H.-T.}\ \bibnamefont
  {Janka}}, \bibinfo {author} {\bibfnamefont {F.}~\bibnamefont {Hanke}},
  \bibinfo {author} {\bibfnamefont {L.}~\bibnamefont {Huedepohl}}, \bibinfo
  {author} {\bibfnamefont {A.}~\bibnamefont {Marek}}, \bibinfo {author}
  {\bibfnamefont {B.}~\bibnamefont {Mueller}}, \ and\ \bibinfo {author}
  {\bibfnamefont {M.}~\bibnamefont {Obergaulinger}},\ }\href
  {http://arxiv.org/abs/1211.1378} {\bibfield  {journal} {\bibinfo  {journal}
  {arXiv:1211.1378 [astro-ph]}\ } (\bibinfo {year} {2012})},\ \bibinfo {note}
  {arXiv: 1211.1378}\BibitemShut {NoStop}%
\bibitem [{\citenamefont {Burrows}(2013)}]{burrows_perspectives_2013}%
  \BibitemOpen
  \bibfield  {author} {\bibinfo {author} {\bibfnamefont {A.}~\bibnamefont
  {Burrows}},\ }\href {\doibase 10.1103/RevModPhys.85.245} {\bibfield
  {journal} {\bibinfo  {journal} {Rev. Mod. Phys.}\ }\textbf {\bibinfo {volume}
  {85}},\ \bibinfo {pages} {245} (\bibinfo {year} {2013})},\ \bibinfo {note}
  {arXiv: 1210.4921}\BibitemShut {NoStop}%
\bibitem [{\citenamefont {Foglizzo}\ \emph {et~al.}(2015)\citenamefont
  {Foglizzo}, \citenamefont {Kazeroni}, \citenamefont {Guilet}, \citenamefont
  {Masset}, \citenamefont {González}, \citenamefont {Krueger}, \citenamefont
  {Novak}, \citenamefont {Oertel}, \citenamefont {Margueron}, \citenamefont
  {Faure}, \citenamefont {Martin}, \citenamefont {Blottiau}, \citenamefont
  {Peres},\ and\ \citenamefont {Durand}}]{foglizzo_explosion_2015}%
  \BibitemOpen
  \bibfield  {author} {\bibinfo {author} {\bibfnamefont {T.}~\bibnamefont
  {Foglizzo}}, \bibinfo {author} {\bibfnamefont {R.}~\bibnamefont {Kazeroni}},
  \bibinfo {author} {\bibfnamefont {J.}~\bibnamefont {Guilet}}, \bibinfo
  {author} {\bibfnamefont {F.}~\bibnamefont {Masset}}, \bibinfo {author}
  {\bibfnamefont {M.}~\bibnamefont {González}}, \bibinfo {author}
  {\bibfnamefont {B.~K.}\ \bibnamefont {Krueger}}, \bibinfo {author}
  {\bibfnamefont {J.}~\bibnamefont {Novak}}, \bibinfo {author} {\bibfnamefont
  {M.}~\bibnamefont {Oertel}}, \bibinfo {author} {\bibfnamefont
  {J.}~\bibnamefont {Margueron}}, \bibinfo {author} {\bibfnamefont
  {J.}~\bibnamefont {Faure}}, \bibinfo {author} {\bibfnamefont
  {N.}~\bibnamefont {Martin}}, \bibinfo {author} {\bibfnamefont
  {P.}~\bibnamefont {Blottiau}}, \bibinfo {author} {\bibfnamefont
  {B.}~\bibnamefont {Peres}}, \ and\ \bibinfo {author} {\bibfnamefont
  {G.}~\bibnamefont {Durand}},\ }\href {\doibase 10.1017/pasa.2015.9}
  {\bibfield  {journal} {\bibinfo  {journal} {Publ. Astron. Soc. Aust.}\
  }\textbf {\bibinfo {volume} {32}},\ \bibinfo {pages} {e009} (\bibinfo {year}
  {2015})},\ \bibinfo {note} {arXiv: 1501.01334}\BibitemShut {NoStop}%
\bibitem [{\citenamefont {Vartanyan}\ \emph
  {et~al.}(2018{\natexlab{b}})\citenamefont {Vartanyan}, \citenamefont
  {Burrows}, \citenamefont {Radice}, \citenamefont {Skinner},\ and\
  \citenamefont {Dolence}}]{vartanyan_successful_2018}%
  \BibitemOpen
  \bibfield  {author} {\bibinfo {author} {\bibfnamefont {D.}~\bibnamefont
  {Vartanyan}}, \bibinfo {author} {\bibfnamefont {A.}~\bibnamefont {Burrows}},
  \bibinfo {author} {\bibfnamefont {D.}~\bibnamefont {Radice}}, \bibinfo
  {author} {\bibfnamefont {A.}~\bibnamefont {Skinner}}, \ and\ \bibinfo
  {author} {\bibfnamefont {J.}~\bibnamefont {Dolence}},\ }\href
  {http://arxiv.org/abs/1809.05106} {\bibfield  {journal} {\bibinfo  {journal}
  {arXiv:1809.05106 [astro-ph]}\ } (\bibinfo {year} {2018}{\natexlab{b}})},\
  \bibinfo {note} {arXiv: 1809.05106}\BibitemShut {NoStop}%
\bibitem [{\citenamefont {Vartanyan}\ \emph {et~al.}(2019)\citenamefont
  {Vartanyan}, \citenamefont {Burrows},\ and\ \citenamefont
  {Radice}}]{vartanyan_temporal_2019}%
  \BibitemOpen
  \bibfield  {author} {\bibinfo {author} {\bibfnamefont {D.}~\bibnamefont
  {Vartanyan}}, \bibinfo {author} {\bibfnamefont {A.}~\bibnamefont {Burrows}},
  \ and\ \bibinfo {author} {\bibfnamefont {D.}~\bibnamefont {Radice}},\ }\href
  {http://arxiv.org/abs/1906.08787} {\bibfield  {journal} {\bibinfo  {journal}
  {arXiv:1906.08787 [astro-ph]}\ } (\bibinfo {year} {2019})},\ \bibinfo {note}
  {arXiv: 1906.08787}\BibitemShut {NoStop}%
\bibitem [{\citenamefont {Nagakura}\ \emph
  {et~al.}(2019{\natexlab{a}})\citenamefont {Nagakura}, \citenamefont
  {Burrows}, \citenamefont {Radice},\ and\ \citenamefont
  {Vartanyan}}]{nagakura_towards_2019}%
  \BibitemOpen
  \bibfield  {author} {\bibinfo {author} {\bibfnamefont {H.}~\bibnamefont
  {Nagakura}}, \bibinfo {author} {\bibfnamefont {A.}~\bibnamefont {Burrows}},
  \bibinfo {author} {\bibfnamefont {D.}~\bibnamefont {Radice}}, \ and\ \bibinfo
  {author} {\bibfnamefont {D.}~\bibnamefont {Vartanyan}},\ }\href {\doibase
  10.1093/mnras/stz2730} {\bibfield  {journal} {\bibinfo  {journal} {Monthly
  Notices of the Royal Astronomical Society}\ }\textbf {\bibinfo {volume}
  {490}},\ \bibinfo {pages} {4622} (\bibinfo {year}
  {2019}{\natexlab{a}})}\BibitemShut {NoStop}%
\bibitem [{\citenamefont {Murphy}\ \emph {et~al.}(2019)\citenamefont {Murphy},
  \citenamefont {Mabanta},\ and\ \citenamefont
  {Dolence}}]{murphy_comparison_2019}%
  \BibitemOpen
  \bibfield  {author} {\bibinfo {author} {\bibfnamefont {J.~W.}\ \bibnamefont
  {Murphy}}, \bibinfo {author} {\bibfnamefont {Q.}~\bibnamefont {Mabanta}}, \
  and\ \bibinfo {author} {\bibfnamefont {J.~C.}\ \bibnamefont {Dolence}},\
  }\href {http://arxiv.org/abs/1904.09444} {\bibfield  {journal} {\bibinfo
  {journal} {arXiv:1904.09444 [astro-ph]}\ } (\bibinfo {year} {2019})},\
  \bibinfo {note} {arXiv: 1904.09444}\BibitemShut {NoStop}%
\bibitem [{\citenamefont {Müller}(2016)}]{muller_status_2016}%
  \BibitemOpen
  \bibfield  {author} {\bibinfo {author} {\bibfnamefont {B.}~\bibnamefont
  {Müller}},\ }\href {\doibase 10.1017/pasa.2016.40} {\bibfield  {journal}
  {\bibinfo  {journal} {Publications of the Astronomical Society of Australia}\
  }\textbf {\bibinfo {volume} {33}} (\bibinfo {year} {2016}),\
  10.1017/pasa.2016.40}\BibitemShut {NoStop}%
\bibitem [{\citenamefont {Müller}\ \emph
  {et~al.}(2018{\natexlab{a}})\citenamefont {Müller}, \citenamefont {Gay},
  \citenamefont {Heger}, \citenamefont {Tauris},\ and\ \citenamefont
  {Sim}}]{muller_multi-d_2018}%
  \BibitemOpen
  \bibfield  {author} {\bibinfo {author} {\bibfnamefont {B.}~\bibnamefont
  {Müller}}, \bibinfo {author} {\bibfnamefont {D.}~\bibnamefont {Gay}},
  \bibinfo {author} {\bibfnamefont {A.}~\bibnamefont {Heger}}, \bibinfo
  {author} {\bibfnamefont {T.}~\bibnamefont {Tauris}}, \ and\ \bibinfo {author}
  {\bibfnamefont {S.~A.}\ \bibnamefont {Sim}},\ }\href {\doibase
  10.1093/mnras/sty1683} {\bibfield  {journal} {\bibinfo  {journal} {Monthly
  Notices of the Royal Astronomical Society}\ }\textbf {\bibinfo {volume}
  {479}},\ \bibinfo {pages} {3675} (\bibinfo {year} {2018}{\natexlab{a}})},\
  \bibinfo {note} {arXiv: 1803.03388}\BibitemShut {NoStop}%
\bibitem [{\citenamefont {Müller}\ \emph
  {et~al.}(2018{\natexlab{b}})\citenamefont {Müller}, \citenamefont {Tauris},
  \citenamefont {Heger}, \citenamefont {Banerjee}, \citenamefont {Qian},
  \citenamefont {Powell}, \citenamefont {Chan}, \citenamefont {Gay},\ and\
  \citenamefont {Langer}}]{muller_three-dimensional_2018}%
  \BibitemOpen
  \bibfield  {author} {\bibinfo {author} {\bibfnamefont {B.}~\bibnamefont
  {Müller}}, \bibinfo {author} {\bibfnamefont {T.~M.}\ \bibnamefont {Tauris}},
  \bibinfo {author} {\bibfnamefont {A.}~\bibnamefont {Heger}}, \bibinfo
  {author} {\bibfnamefont {P.}~\bibnamefont {Banerjee}}, \bibinfo {author}
  {\bibfnamefont {Y.-Z.}\ \bibnamefont {Qian}}, \bibinfo {author}
  {\bibfnamefont {J.}~\bibnamefont {Powell}}, \bibinfo {author} {\bibfnamefont
  {C.}~\bibnamefont {Chan}}, \bibinfo {author} {\bibfnamefont {D.~W.}\
  \bibnamefont {Gay}}, \ and\ \bibinfo {author} {\bibfnamefont
  {N.}~\bibnamefont {Langer}},\ }\href {http://arxiv.org/abs/1811.05483}
  {\bibfield  {journal} {\bibinfo  {journal} {arXiv:1811.05483 [astro-ph]}\ }
  (\bibinfo {year} {2018}{\natexlab{b}})},\ \bibinfo {note} {arXiv:
  1811.05483}\BibitemShut {NoStop}%
\bibitem [{\citenamefont {Glas}\ \emph {et~al.}(2019)\citenamefont {Glas},
  \citenamefont {Just}, \citenamefont {Janka},\ and\ \citenamefont
  {Obergaulinger}}]{glas_three-dimensional_2019}%
  \BibitemOpen
  \bibfield  {author} {\bibinfo {author} {\bibfnamefont {R.}~\bibnamefont
  {Glas}}, \bibinfo {author} {\bibfnamefont {O.}~\bibnamefont {Just}}, \bibinfo
  {author} {\bibfnamefont {H.-T.}\ \bibnamefont {Janka}}, \ and\ \bibinfo
  {author} {\bibfnamefont {M.}~\bibnamefont {Obergaulinger}},\ }\href {\doibase
  10.3847/1538-4357/ab0423} {\bibfield  {journal} {\bibinfo  {journal} {ApJ}\
  }\textbf {\bibinfo {volume} {873}},\ \bibinfo {pages} {45} (\bibinfo {year}
  {2019})},\ \bibinfo {note} {arXiv: 1809.10146}\BibitemShut {NoStop}%
\bibitem [{\citenamefont {Radice}\ \emph {et~al.}(2018)\citenamefont {Radice},
  \citenamefont {Morozova}, \citenamefont {Burrows}, \citenamefont
  {Vartanyan},\ and\ \citenamefont {Nagakura}}]{radice_characterizing_2018}%
  \BibitemOpen
  \bibfield  {author} {\bibinfo {author} {\bibfnamefont {D.}~\bibnamefont
  {Radice}}, \bibinfo {author} {\bibfnamefont {V.}~\bibnamefont {Morozova}},
  \bibinfo {author} {\bibfnamefont {A.}~\bibnamefont {Burrows}}, \bibinfo
  {author} {\bibfnamefont {D.}~\bibnamefont {Vartanyan}}, \ and\ \bibinfo
  {author} {\bibfnamefont {H.}~\bibnamefont {Nagakura}},\ }\href
  {http://arxiv.org/abs/1812.07703} {\bibfield  {journal} {\bibinfo  {journal}
  {arXiv:1812.07703 [astro-ph, physics:gr-qc]}\ } (\bibinfo {year} {2018})},\
  \bibinfo {note} {arXiv: 1812.07703}\BibitemShut {NoStop}%
\bibitem [{\citenamefont {Powell}\ and\ \citenamefont
  {Müller}(2018)}]{powell_gravitational_2018}%
  \BibitemOpen
  \bibfield  {author} {\bibinfo {author} {\bibfnamefont {J.}~\bibnamefont
  {Powell}}\ and\ \bibinfo {author} {\bibfnamefont {B.}~\bibnamefont
  {Müller}},\ }\href {http://arxiv.org/abs/1812.05738} {\bibfield  {journal}
  {\bibinfo  {journal} {arXiv:1812.05738 [astro-ph]}\ } (\bibinfo {year}
  {2018})},\ \bibinfo {note} {arXiv: 1812.05738}\BibitemShut {NoStop}%
\bibitem [{\citenamefont {O'Connor}\ and\ \citenamefont
  {Couch}(2018)}]{oconnor_exploring_2018}%
  \BibitemOpen
  \bibfield  {author} {\bibinfo {author} {\bibfnamefont {E.}~\bibnamefont
  {O'Connor}}\ and\ \bibinfo {author} {\bibfnamefont {S.}~\bibnamefont
  {Couch}},\ }\href {http://arxiv.org/abs/1807.07579} {\bibfield  {journal}
  {\bibinfo  {journal} {arXiv:1807.07579 [astro-ph]}\ } (\bibinfo {year}
  {2018})},\ \bibinfo {note} {arXiv: 1807.07579}\BibitemShut {NoStop}%
\bibitem [{\citenamefont {Cabezón}\ \emph {et~al.}(2018)\citenamefont
  {Cabezón}, \citenamefont {Pan}, \citenamefont {Liebendörfer}, \citenamefont
  {Kuroda}, \citenamefont {Ebinger}, \citenamefont {Heinimann}, \citenamefont
  {Thielemann},\ and\ \citenamefont {Perego}}]{cabezon_core-collapse_2018}%
  \BibitemOpen
  \bibfield  {author} {\bibinfo {author} {\bibfnamefont {R.~M.}\ \bibnamefont
  {Cabezón}}, \bibinfo {author} {\bibfnamefont {K.-C.}\ \bibnamefont {Pan}},
  \bibinfo {author} {\bibfnamefont {M.}~\bibnamefont {Liebendörfer}}, \bibinfo
  {author} {\bibfnamefont {T.}~\bibnamefont {Kuroda}}, \bibinfo {author}
  {\bibfnamefont {K.}~\bibnamefont {Ebinger}}, \bibinfo {author} {\bibfnamefont
  {O.}~\bibnamefont {Heinimann}}, \bibinfo {author} {\bibfnamefont {F.-K.}\
  \bibnamefont {Thielemann}}, \ and\ \bibinfo {author} {\bibfnamefont
  {A.}~\bibnamefont {Perego}},\ }\href {http://arxiv.org/abs/1806.09184}
  {\bibfield  {journal} {\bibinfo  {journal} {arXiv:1806.09184 [astro-ph]}\ }
  (\bibinfo {year} {2018})},\ \bibinfo {note} {arXiv: 1806.09184}\BibitemShut
  {NoStop}%
\bibitem [{\citenamefont {Nagakura}\ \emph
  {et~al.}(2019{\natexlab{b}})\citenamefont {Nagakura}, \citenamefont
  {Sumiyoshi},\ and\ \citenamefont {Yamada}}]{nagakura_three-dimensional_2019}%
  \BibitemOpen
  \bibfield  {author} {\bibinfo {author} {\bibfnamefont {H.}~\bibnamefont
  {Nagakura}}, \bibinfo {author} {\bibfnamefont {K.}~\bibnamefont {Sumiyoshi}},
  \ and\ \bibinfo {author} {\bibfnamefont {S.}~\bibnamefont {Yamada}},\ }\href
  {http://arxiv.org/abs/1906.10143} {\bibfield  {journal} {\bibinfo  {journal}
  {arXiv:1906.10143 [astro-ph]}\ } (\bibinfo {year} {2019}{\natexlab{b}})},\
  \bibinfo {note} {arXiv: 1906.10143}\BibitemShut {NoStop}%
\bibitem [{\citenamefont {Thorne}(1981)}]{thorne_relativistic_1981}%
  \BibitemOpen
  \bibfield  {author} {\bibinfo {author} {\bibfnamefont {K.~S.}\ \bibnamefont
  {Thorne}},\ }\href {\doibase 10.1093/mnras/194.2.439} {\bibfield  {journal}
  {\bibinfo  {journal} {Monthly Notices of the Royal Astronomical Society}\
  }\textbf {\bibinfo {volume} {194}},\ \bibinfo {pages} {439} (\bibinfo {year}
  {1981})}\BibitemShut {NoStop}%
\bibitem [{\citenamefont {Shibata}\ \emph {et~al.}(2011)\citenamefont
  {Shibata}, \citenamefont {Kiuchi}, \citenamefont {Sekiguchi},\ and\
  \citenamefont {Suwa}}]{shibata_truncated_2011}%
  \BibitemOpen
  \bibfield  {author} {\bibinfo {author} {\bibfnamefont {M.}~\bibnamefont
  {Shibata}}, \bibinfo {author} {\bibfnamefont {K.}~\bibnamefont {Kiuchi}},
  \bibinfo {author} {\bibfnamefont {Y.-i.}\ \bibnamefont {Sekiguchi}}, \ and\
  \bibinfo {author} {\bibfnamefont {Y.}~\bibnamefont {Suwa}},\ }\href {\doibase
  10.1143/PTP.125.1255} {\bibfield  {journal} {\bibinfo  {journal} {Progress of
  Theoretical Physics}\ }\textbf {\bibinfo {volume} {125}},\ \bibinfo {pages}
  {1255} (\bibinfo {year} {2011})},\ \bibinfo {note} {arXiv:
  1104.3937}\BibitemShut {NoStop}%
\bibitem [{\citenamefont {Bollig}\ \emph {et~al.}(2017)\citenamefont {Bollig},
  \citenamefont {Janka}, \citenamefont {Lohs}, \citenamefont
  {Martínez-Pinedo}, \citenamefont {Horowitz},\ and\ \citenamefont
  {Melson}}]{bollig_muon_2017}%
  \BibitemOpen
  \bibfield  {author} {\bibinfo {author} {\bibfnamefont {R.}~\bibnamefont
  {Bollig}}, \bibinfo {author} {\bibfnamefont {H.-T.}\ \bibnamefont {Janka}},
  \bibinfo {author} {\bibfnamefont {A.}~\bibnamefont {Lohs}}, \bibinfo {author}
  {\bibfnamefont {G.}~\bibnamefont {Martínez-Pinedo}}, \bibinfo {author}
  {\bibfnamefont {C.}~\bibnamefont {Horowitz}}, \ and\ \bibinfo {author}
  {\bibfnamefont {T.}~\bibnamefont {Melson}},\ }\href {\doibase
  10.1103/PhysRevLett.119.242702} {\bibfield  {journal} {\bibinfo  {journal}
  {Phys. Rev. Lett.}\ }\textbf {\bibinfo {volume} {119}},\ \bibinfo {pages}
  {242702} (\bibinfo {year} {2017})},\ \bibinfo {note} {publisher: American
  Physical Society}\BibitemShut {NoStop}%
\bibitem [{\citenamefont {Guo}\ \emph {et~al.}(2020)\citenamefont {Guo},
  \citenamefont {Martínez-Pinedo}, \citenamefont {Lohs},\ and\ \citenamefont
  {Fischer}}]{guo_charged-current_2020}%
  \BibitemOpen
  \bibfield  {author} {\bibinfo {author} {\bibfnamefont {G.}~\bibnamefont
  {Guo}}, \bibinfo {author} {\bibfnamefont {G.}~\bibnamefont
  {Martínez-Pinedo}}, \bibinfo {author} {\bibfnamefont {A.}~\bibnamefont
  {Lohs}}, \ and\ \bibinfo {author} {\bibfnamefont {T.}~\bibnamefont
  {Fischer}},\ }\href {http://arxiv.org/abs/2006.12051} {\bibfield  {journal}
  {\bibinfo  {journal} {arXiv:2006.12051 [astro-ph, physics:hep-ph]}\ }
  (\bibinfo {year} {2020})},\ \bibinfo {note} {arXiv: 2006.12051}\BibitemShut
  {NoStop}%
\bibitem [{\citenamefont {Fischer}\ \emph {et~al.}(2020)\citenamefont
  {Fischer}, \citenamefont {Typel}, \citenamefont {Röpke}, \citenamefont
  {Bastian},\ and\ \citenamefont {Martínez-Pinedo}}]{fischer_medium_2020}%
  \BibitemOpen
  \bibfield  {author} {\bibinfo {author} {\bibfnamefont {T.}~\bibnamefont
  {Fischer}}, \bibinfo {author} {\bibfnamefont {S.}~\bibnamefont {Typel}},
  \bibinfo {author} {\bibfnamefont {G.}~\bibnamefont {Röpke}}, \bibinfo
  {author} {\bibfnamefont {N.-U.~F.}\ \bibnamefont {Bastian}}, \ and\ \bibinfo
  {author} {\bibfnamefont {G.}~\bibnamefont {Martínez-Pinedo}},\ }\href
  {http://arxiv.org/abs/2008.13608} {\bibfield  {journal} {\bibinfo  {journal}
  {arXiv:2008.13608 [astro-ph, physics:nucl-th]}\ } (\bibinfo {year} {2020})},\
  \bibinfo {note} {arXiv: 2008.13608}\BibitemShut {NoStop}%
\bibitem [{\citenamefont {O'Connor}(2015)}]{oconnor_open-source_2015}%
  \BibitemOpen
  \bibfield  {author} {\bibinfo {author} {\bibfnamefont {E.}~\bibnamefont
  {O'Connor}},\ }\href {\doibase 10.1088/0067-0049/219/2/24} {\bibfield
  {journal} {\bibinfo  {journal} {ApJS}\ }\textbf {\bibinfo {volume} {219}},\
  \bibinfo {pages} {24} (\bibinfo {year} {2015})},\ \bibinfo {note} {arXiv:
  1411.7058}\BibitemShut {NoStop}%
\bibitem [{\citenamefont {Hannestad}\ and\ \citenamefont
  {Raffelt}(1998)}]{hannestad_supernova_1998}%
  \BibitemOpen
  \bibfield  {author} {\bibinfo {author} {\bibfnamefont {S.}~\bibnamefont
  {Hannestad}}\ and\ \bibinfo {author} {\bibfnamefont {G.}~\bibnamefont
  {Raffelt}},\ }\href {\doibase 10.1086/306303} {\bibfield  {journal} {\bibinfo
   {journal} {The Astrophysical Journal}\ }\textbf {\bibinfo {volume} {507}},\
  \bibinfo {pages} {339} (\bibinfo {year} {1998})},\ \bibinfo {note} {arXiv:
  astro-ph/9711132}\BibitemShut {NoStop}%
\bibitem [{\citenamefont {{Guo}}\ and\ \citenamefont
  {{Mart{\'\i}nez-Pinedo}}(2019)}]{guo_chiral_2019}%
  \BibitemOpen
  \bibfield  {author} {\bibinfo {author} {\bibfnamefont {G.}~\bibnamefont
  {{Guo}}}\ and\ \bibinfo {author} {\bibfnamefont {G.}~\bibnamefont
  {{Mart{\'\i}nez-Pinedo}}},\ }\href {\doibase 10.3847/1538-4357/ab536d}
  {\bibfield  {journal} {\bibinfo  {journal} {\apj}\ }\textbf {\bibinfo
  {volume} {887}},\ \bibinfo {eid} {58} (\bibinfo {year} {2019})},\ \Eprint
  {http://arxiv.org/abs/1905.13634} {arXiv:1905.13634 [astro-ph.HE]}
  \BibitemShut {NoStop}%
\bibitem [{\citenamefont {Burrows}\ \emph {et~al.}(2006)\citenamefont
  {Burrows}, \citenamefont {Reddy},\ and\ \citenamefont
  {Thompson}}]{burrows_neutrino_2006}%
  \BibitemOpen
  \bibfield  {author} {\bibinfo {author} {\bibfnamefont {A.}~\bibnamefont
  {Burrows}}, \bibinfo {author} {\bibfnamefont {S.}~\bibnamefont {Reddy}}, \
  and\ \bibinfo {author} {\bibfnamefont {T.~A.}\ \bibnamefont {Thompson}},\
  }\href {\doibase 10.1016/j.nuclphysa.2004.06.012} {\bibfield  {journal}
  {\bibinfo  {journal} {Nuclear Physics A}\ }\textbf {\bibinfo {volume}
  {777}},\ \bibinfo {pages} {356} (\bibinfo {year} {2006})},\ \bibinfo {note}
  {arXiv: astro-ph/0404432}\BibitemShut {NoStop}%
\bibitem [{\citenamefont {Bruenn}(1985)}]{bruenn_stellar_1985}%
  \BibitemOpen
  \bibfield  {author} {\bibinfo {author} {\bibfnamefont {S.~W.}\ \bibnamefont
  {Bruenn}},\ }\href {\doibase 10.1086/191056} {\bibfield  {journal} {\bibinfo
  {journal} {The Astrophysical Journal Supplement Series}\ }\textbf {\bibinfo
  {volume} {58}},\ \bibinfo {pages} {771} (\bibinfo {year} {1985})}\BibitemShut
  {NoStop}%
\bibitem [{\citenamefont {O'Connor}\ and\ \citenamefont
  {Ott}(2010)}]{oconnor_new_2010}%
  \BibitemOpen
  \bibfield  {author} {\bibinfo {author} {\bibfnamefont {E.}~\bibnamefont
  {O'Connor}}\ and\ \bibinfo {author} {\bibfnamefont {C.~D.}\ \bibnamefont
  {Ott}},\ }\href {\doibase 10.1088/0264-9381/27/11/114103} {\bibfield
  {journal} {\bibinfo  {journal} {Class. Quantum Grav.}\ }\textbf {\bibinfo
  {volume} {27}},\ \bibinfo {pages} {114103} (\bibinfo {year} {2010})},\
  \bibinfo {note} {arXiv: 0912.2393}\BibitemShut {NoStop}%
\bibitem [{\citenamefont {Cardall}\ \emph {et~al.}(2013)\citenamefont
  {Cardall}, \citenamefont {Endeve},\ and\ \citenamefont
  {Mezzacappa}}]{cardall_conservative_2013}%
  \BibitemOpen
  \bibfield  {author} {\bibinfo {author} {\bibfnamefont {C.~Y.}\ \bibnamefont
  {Cardall}}, \bibinfo {author} {\bibfnamefont {E.}~\bibnamefont {Endeve}}, \
  and\ \bibinfo {author} {\bibfnamefont {A.}~\bibnamefont {Mezzacappa}},\
  }\href {\doibase 10.1103/PhysRevD.87.103004} {\bibfield  {journal} {\bibinfo
  {journal} {Physical Review D}\ }\textbf {\bibinfo {volume} {87}} (\bibinfo
  {year} {2013}),\ 10.1103/PhysRevD.87.103004}\BibitemShut {NoStop}%
\bibitem [{\citenamefont {Horowitz}(2002)}]{horowitz_weak_2002}%
  \BibitemOpen
  \bibfield  {author} {\bibinfo {author} {\bibfnamefont {C.~J.}\ \bibnamefont
  {Horowitz}},\ }\href {\doibase 10.1103/PhysRevD.65.043001} {\bibfield
  {journal} {\bibinfo  {journal} {Physical Review D}\ }\textbf {\bibinfo
  {volume} {65}} (\bibinfo {year} {2002}),\ 10.1103/PhysRevD.65.043001},\
  \bibinfo {note} {arXiv: astro-ph/0109209}\BibitemShut {NoStop}%
\bibitem [{\citenamefont {Horowitz}(1997)}]{horowitz_neutrino_1997}%
  \BibitemOpen
  \bibfield  {author} {\bibinfo {author} {\bibfnamefont {C.~J.}\ \bibnamefont
  {Horowitz}},\ }\href {\doibase 10.1103/PhysRevD.55.4577} {\bibfield
  {journal} {\bibinfo  {journal} {Physical Review D}\ }\textbf {\bibinfo
  {volume} {55}},\ \bibinfo {pages} {4577} (\bibinfo {year} {1997})},\ \bibinfo
  {note} {arXiv: astro-ph/9603138}\BibitemShut {NoStop}%
\bibitem [{\citenamefont {Entem}\ \emph {et~al.}(2017)\citenamefont {Entem},
  \citenamefont {Machleidt},\ and\ \citenamefont
  {Nosyk}}]{entem_high-quality_2017}%
  \BibitemOpen
  \bibfield  {author} {\bibinfo {author} {\bibfnamefont {D.~R.}\ \bibnamefont
  {Entem}}, \bibinfo {author} {\bibfnamefont {R.}~\bibnamefont {Machleidt}}, \
  and\ \bibinfo {author} {\bibfnamefont {Y.}~\bibnamefont {Nosyk}},\ }\href
  {\doibase 10.1103/PhysRevC.96.024004} {\bibfield  {journal} {\bibinfo
  {journal} {Phys. Rev. C}\ }\textbf {\bibinfo {volume} {96}},\ \bibinfo
  {pages} {024004} (\bibinfo {year} {2017})},\ \bibinfo {note} {publisher:
  American Physical Society}\BibitemShut {NoStop}%
\bibitem [{\citenamefont {Bartl}\ \emph {et~al.}(2014)\citenamefont {Bartl},
  \citenamefont {Pethick},\ and\ \citenamefont
  {Schwenk}}]{bartl_supernova_2014}%
  \BibitemOpen
  \bibfield  {author} {\bibinfo {author} {\bibfnamefont {A.}~\bibnamefont
  {Bartl}}, \bibinfo {author} {\bibfnamefont {C.~J.}\ \bibnamefont {Pethick}},
  \ and\ \bibinfo {author} {\bibfnamefont {A.}~\bibnamefont {Schwenk}},\ }\href
  {\doibase 10.1103/PhysRevLett.113.081101} {\bibfield  {journal} {\bibinfo
  {journal} {Phys. Rev. Lett.}\ }\textbf {\bibinfo {volume} {113}},\ \bibinfo
  {pages} {081101} (\bibinfo {year} {2014})},\ \bibinfo {note} {arXiv:
  1403.4114}\BibitemShut {NoStop}%
\bibitem [{\citenamefont {Bartl}\ \emph {et~al.}(2016)\citenamefont {Bartl},
  \citenamefont {Bollig}, \citenamefont {Janka},\ and\ \citenamefont
  {Schwenk}}]{bartl_impact_2016}%
  \BibitemOpen
  \bibfield  {author} {\bibinfo {author} {\bibfnamefont {A.}~\bibnamefont
  {Bartl}}, \bibinfo {author} {\bibfnamefont {R.}~\bibnamefont {Bollig}},
  \bibinfo {author} {\bibfnamefont {H.-T.}\ \bibnamefont {Janka}}, \ and\
  \bibinfo {author} {\bibfnamefont {A.}~\bibnamefont {Schwenk}},\ }\href
  {\doibase 10.1103/PhysRevD.94.083009} {\bibfield  {journal} {\bibinfo
  {journal} {Physical Review D}\ }\textbf {\bibinfo {volume} {94}} (\bibinfo
  {year} {2016}),\ 10.1103/PhysRevD.94.083009},\ \bibinfo {note} {arXiv:
  1608.05037}\BibitemShut {NoStop}%
\bibitem [{\citenamefont {Woosley}\ and\ \citenamefont
  {Heger}(2007)}]{woosley_nucleosynthesis_2007}%
  \BibitemOpen
  \bibfield  {author} {\bibinfo {author} {\bibfnamefont {S.~E.}\ \bibnamefont
  {Woosley}}\ and\ \bibinfo {author} {\bibfnamefont {A.}~\bibnamefont
  {Heger}},\ }\href {\doibase 10.1016/j.physrep.2007.02.009} {\bibfield
  {journal} {\bibinfo  {journal} {Physics Reports}\ }\bibinfo {series} {The
  {Hans} {Bethe} {Centennial} {Volume} 1906-2006},\ \textbf {\bibinfo {volume}
  {442}},\ \bibinfo {pages} {269} (\bibinfo {year} {2007})}\BibitemShut
  {NoStop}%
\bibitem [{\citenamefont {Heger}(2018)}]{heger:pc18}%
  \BibitemOpen
  \bibfield  {author} {\bibinfo {author} {\bibfnamefont {A.}~\bibnamefont
  {Heger}},\ }\href@noop {} {}\bibinfo {howpublished} {{Private Communication}}
  (\bibinfo {year} {2018})\BibitemShut {NoStop}%
\bibitem [{\citenamefont {Just}\ \emph {et~al.}(2018)\citenamefont {Just},
  \citenamefont {Bollig}, \citenamefont {Janka}, \citenamefont {Obergaulinger},
  \citenamefont {Glas},\ and\ \citenamefont
  {Nagataki}}]{just_core-collapse_2018}%
  \BibitemOpen
  \bibfield  {author} {\bibinfo {author} {\bibfnamefont {O.}~\bibnamefont
  {Just}}, \bibinfo {author} {\bibfnamefont {R.}~\bibnamefont {Bollig}},
  \bibinfo {author} {\bibfnamefont {H.-T.}\ \bibnamefont {Janka}}, \bibinfo
  {author} {\bibfnamefont {M.}~\bibnamefont {Obergaulinger}}, \bibinfo {author}
  {\bibfnamefont {R.}~\bibnamefont {Glas}}, \ and\ \bibinfo {author}
  {\bibfnamefont {S.}~\bibnamefont {Nagataki}},\ }\href {\doibase
  10.1093/mnras/sty2578} {\bibfield  {journal} {\bibinfo  {journal} {Monthly
  Notices of the Royal Astronomical Society}\ }\textbf {\bibinfo {volume}
  {481}},\ \bibinfo {pages} {4786} (\bibinfo {year} {2018})}\BibitemShut
  {NoStop}%
\bibitem [{\citenamefont {Melson}\ \emph
  {et~al.}(2015{\natexlab{a}})\citenamefont {Melson}, \citenamefont {Janka},\
  and\ \citenamefont {Marek}}]{melson_neutrino-driven_2015}%
  \BibitemOpen
  \bibfield  {author} {\bibinfo {author} {\bibfnamefont {T.}~\bibnamefont
  {Melson}}, \bibinfo {author} {\bibfnamefont {H.-T.}\ \bibnamefont {Janka}}, \
  and\ \bibinfo {author} {\bibfnamefont {A.}~\bibnamefont {Marek}},\ }\href
  {\doibase 10.1088/2041-8205/801/2/L24} {\bibfield  {journal} {\bibinfo
  {journal} {ApJ}\ }\textbf {\bibinfo {volume} {801}},\ \bibinfo {pages} {L24}
  (\bibinfo {year} {2015}{\natexlab{a}})},\ \bibinfo {note} {arXiv:
  1501.01961}\BibitemShut {NoStop}%
\bibitem [{\citenamefont {{M{\"u}ller}}\ \emph {et~al.}(2013)\citenamefont
  {{M{\"u}ller}}, \citenamefont {{Janka}},\ and\ \citenamefont
  {{Marek}}}]{mueller_gravitational:13}%
  \BibitemOpen
  \bibfield  {author} {\bibinfo {author} {\bibfnamefont {B.}~\bibnamefont
  {{M{\"u}ller}}}, \bibinfo {author} {\bibfnamefont {H.-T.}\ \bibnamefont
  {{Janka}}}, \ and\ \bibinfo {author} {\bibfnamefont {A.}~\bibnamefont
  {{Marek}}},\ }\href {\doibase 10.1088/0004-637X/766/1/43} {\bibfield
  {journal} {\bibinfo  {journal} {\apj}\ }\textbf {\bibinfo {volume} {766}},\
  \bibinfo {eid} {43} (\bibinfo {year} {2013})},\ \Eprint
  {http://arxiv.org/abs/1210.6984} {arXiv:1210.6984 [astro-ph.SR]} \BibitemShut
  {NoStop}%
\bibitem [{\citenamefont {Radice}\ \emph {et~al.}(2017)\citenamefont {Radice},
  \citenamefont {Burrows}, \citenamefont {Vartanyan}, \citenamefont {Skinner},\
  and\ \citenamefont {Dolence}}]{radice_electron-capture_2017}%
  \BibitemOpen
  \bibfield  {author} {\bibinfo {author} {\bibfnamefont {D.}~\bibnamefont
  {Radice}}, \bibinfo {author} {\bibfnamefont {A.}~\bibnamefont {Burrows}},
  \bibinfo {author} {\bibfnamefont {D.}~\bibnamefont {Vartanyan}}, \bibinfo
  {author} {\bibfnamefont {M.~A.}\ \bibnamefont {Skinner}}, \ and\ \bibinfo
  {author} {\bibfnamefont {J.~C.}\ \bibnamefont {Dolence}},\ }\href {\doibase
  10.3847/1538-4357/aa92c5} {\bibfield  {journal} {\bibinfo  {journal} {ApJ}\
  }\textbf {\bibinfo {volume} {850}},\ \bibinfo {pages} {43} (\bibinfo {year}
  {2017})},\ \bibinfo {note} {arXiv: 1702.03927}\BibitemShut {NoStop}%
\bibitem [{\citenamefont {Steiner}\ \emph {et~al.}(2013)\citenamefont
  {Steiner}, \citenamefont {Hempel},\ and\ \citenamefont
  {Fischer}}]{steiner_core-collapse_2013}%
  \BibitemOpen
  \bibfield  {author} {\bibinfo {author} {\bibfnamefont {A.~W.}\ \bibnamefont
  {Steiner}}, \bibinfo {author} {\bibfnamefont {M.}~\bibnamefont {Hempel}}, \
  and\ \bibinfo {author} {\bibfnamefont {T.}~\bibnamefont {Fischer}},\ }\href
  {\doibase 10.1088/0004-637X/774/1/17} {\bibfield  {journal} {\bibinfo
  {journal} {The Astrophysical Journal}\ }\textbf {\bibinfo {volume} {774}},\
  \bibinfo {pages} {17} (\bibinfo {year} {2013})}\BibitemShut {NoStop}%
\bibitem [{\citenamefont {Melson}\ \emph
  {et~al.}(2015{\natexlab{b}})\citenamefont {Melson}, \citenamefont {Janka},
  \citenamefont {Bollig}, \citenamefont {Hanke}, \citenamefont {Marek},\ and\
  \citenamefont {Müller}}]{melson_neutrino-driven_2015-1}%
  \BibitemOpen
  \bibfield  {author} {\bibinfo {author} {\bibfnamefont {T.}~\bibnamefont
  {Melson}}, \bibinfo {author} {\bibfnamefont {H.-T.}\ \bibnamefont {Janka}},
  \bibinfo {author} {\bibfnamefont {R.}~\bibnamefont {Bollig}}, \bibinfo
  {author} {\bibfnamefont {F.}~\bibnamefont {Hanke}}, \bibinfo {author}
  {\bibfnamefont {A.}~\bibnamefont {Marek}}, \ and\ \bibinfo {author}
  {\bibfnamefont {B.}~\bibnamefont {Müller}},\ }\href {\doibase
  10.1088/2041-8205/808/2/L42} {\bibfield  {journal} {\bibinfo  {journal} {The
  Astrophysical Journal Letters}\ }\textbf {\bibinfo {volume} {808}},\ \bibinfo
  {pages} {L42} (\bibinfo {year} {2015}{\natexlab{b}})}\BibitemShut {NoStop}%
\bibitem [{\citenamefont {O'Connor}(2017)}]{oconnor_core-collapse_2017}%
  \BibitemOpen
  \bibfield  {author} {\bibinfo {author} {\bibfnamefont {E.}~\bibnamefont
  {O'Connor}},\ }in\ \href {\doibase 10.1007/978-3-319-21846-5_129} {\emph
  {\bibinfo {booktitle} {Handbook of Supernovae}}},\ \bibinfo {editor} {edited
  by\ \bibinfo {editor} {\bibfnamefont {A.~W.}\ \bibnamefont {Alsabti}}\ and\
  \bibinfo {editor} {\bibfnamefont {P.}~\bibnamefont {Murdin}}}\ (\bibinfo
  {publisher} {Springer International Publishing},\ \bibinfo {address} {Cham},\
  \bibinfo {year} {2017})\ pp.\ \bibinfo {pages} {1555--1572}\BibitemShut
  {NoStop}%
\bibitem [{\citenamefont {{Timmes}}\ and\ \citenamefont
  {{Arnett}}(1999)}]{timmes:99}%
  \BibitemOpen
  \bibfield  {author} {\bibinfo {author} {\bibfnamefont {F.~X.}\ \bibnamefont
  {{Timmes}}}\ and\ \bibinfo {author} {\bibfnamefont {D.}~\bibnamefont
  {{Arnett}}},\ }\href {\doibase 10.1086/313271} {\bibfield  {journal}
  {\bibinfo  {journal} {ApJS}\ }\textbf {\bibinfo {volume} {125}},\ \bibinfo
  {pages} {277} (\bibinfo {year} {1999})}\BibitemShut {NoStop}%
\end{thebibliography}%

\end{document}